\documentclass[aps,pra,reprint,superscriptaddress,showpacs]{revtex4-1}

\usepackage{graphics}
\usepackage{bm}

\begin{document}

\title{Analytic description of atomic interaction at ultracold temperatures II: 
	Scattering around a magnetic Feshbach resonance}

\author{Bo Gao}
\email[]{bo.gao@utoledo.edu}
\homepage[]{http://bgaowww.physics.utoledo.edu}
\affiliation{Institute for Theoretical Atomic, Molecular and Optical Physics (ITAMP),
Harvard-Smithsonian Center for Astrophysics, 
Cambridge, Massachusetts 02138, USA}
\affiliation{Department of Physics and Astronomy,
Mailstop 111,
University of Toledo,
Toledo, Ohio 43606, 
USA}

\date{\today}

\begin{abstract}

Starting from a multichannel quantum-defect theory, we derive 
analytic descriptions of a magnetic Feshbach resonance in an arbitrary 
partial wave $l$, and the atomic interactions around it.
An analytic formula, applicable to both broad and narrow resonances
of arbitrary $l$, is presented for ultracold atomic scattering 
around a Feshbach resonance.
Other related issues addressed include (a) the parametrization of a magnetic
Feshbach resonance of arbitrary $l$, (b) rigorous definitions of 
``broad'' and ``narrow'' resonances of arbitrary $l$ and their different
scattering characteristics, and (c) the tuning of the effective range 
and the generalized effective range by a magnetic field.

\end{abstract}

\pacs{34.10.+x,34.50.Cx,33.15.-e,03.75.Nt}

\maketitle

\section{Introduction}

Analytic descriptions of two-body interactions are highly desirable
if any systematic understanding of quantum few-body and quantum
many-body systems is to be expected or achieved.
The best-known example may be the Gross-Pitaevskii
theory of identical bosons \cite{[{See, e.g, }][]dal99}, 
with its simplicity and generality
depending intimately on our ability to parametrize the low-energy
two-body interaction using the effective range
theory (ERT) \cite{sch47,bla49,bet49}.
The same is true for quantum few-body theories  
in the universal regime (see, e.g., Refs.~\cite{bra06,gre10}).

In a companion paper \cite{gao09a}, referred to hereafter as paper I, 
we have discussed the limitations of the standard 
ERT \cite{sch47,bla49,bet49} in describing atomic interactions
at low temperatures, and how such limitations are overcome
using expansions derived from the quantum-defect theory (QDT)
for $-1/r^6$ type of long-range potentials \cite{gao98b,gao01,gao08a}.
The focus was on the case of a single channel,
both out of the necessity of theoretical development,
but also to provide a set of single-channel universal behaviors
that will serve as benchmarks for understanding
other types of behaviors.

This article extends this discussion to atomic interaction
around a magnetic Feshbach resonance \cite{tie93,koh06,chi10}
in an arbitrary partial wave $l$. 
It is a nontrivial extension with considerable new
physics as a Feshbach resonance is necessarily a 
multichannel phenomena \cite{tie93,koh06,chi10},
for which few analytic results have been derived in
any general context.
The theory includes the parametrization of the resonance,
the rigorous definitions of ``broad'' and ``narrow'' 
resonances \cite{sto05,sim05,koh06,chi10},
and an analytic description of the atomic scattering properties 
around them.
Such understandings, which have been mostly limited
to the $s$ wave \cite{koh06,chi10},
are not only of interest by themselves,
they are also prerequisites for understanding
atomic interaction in an optical lattice 
\cite{[{See, e.g., }][{ and references therein.}]hal10}, 
and behaviors of quantum few-atom and many-atom systems
around a Feshbach resonance. For nonzero partial waves,
the theory here is a necessity as ERT fails \cite{gao98b,che07,gao09a,zha10}.
Even for the $s$ wave, it offers much improved analytic description, especially
for narrow resonances around which the energy dependence of the
scattering amplitude can become so significant that it has to
be incorporated into the corresponding few-body 
\cite{[{See, e.g., }][{ and Ref.~\cite{sto05}.}]pet04b}
and many-body theories (See, e.g., Refs.~\cite{fu03b,szy05,col07,zin09}).

There are three main steps in developing an analytic description of 
a magnetic Feshbach resonance.
The first is the reduction of the underlying multichannel problem, 
as formulated in a multichannel quantum-defect theory (MQDT) 
of Ref.~\cite{gao05a} to an effective single channel problem. 
The second is an efficient parametrization of
a magnetic Feshbach resonance.
The third is to apply the theory of paper I  \cite{gao09a} 
to obtain the desired results such as the scattering properties 
around the threshold, to be addressed in this article.

The paper is organized as follows. The reduction to an effective
single-channel problem is carried out in Sec.~\ref{sec:reduce}. 
The parametrization of a magnetic Feshbach resonance is addressed
in Sec.~\ref{sec:para}. In particular, 
we derive in Sec.~\ref{sec:galB} the magnetic-field 
dependence of scattering lengths and the generalized scattering lengths
introduced in paper I \cite{gao09a}. We show in this section that regardless of $l$,
the scattering length, or the generalized scattering length for $l\ge 2$,
can be parametrized around a magnetic Feshbach resonance 
in a similar fashion as the $s$ wave scattering length \cite{moe95,koh06,chi10}.
The parametrization is further developed
in Sec.~\ref{sec:scaledpara} in terms of scaled parameters. 
It leads not only to more concise analytic formulas, 
but more importantly, to rigorous definitions of 
``broad'' and ``narrow'' Feshbach resonances of arbitrary $l$.
In Sec.~\ref{sec:scatteringB}, we present the
QDT expansion \cite{gao09a} that provides an analytic description of
ultracold scattering around a magnetic Feshbach resonance of
arbitrary $l$. 
As sample applications of the QDT expansion,
Sec.~\ref{sec:apps} presents and discusses the generalized
effective range expansion \cite{gao09a} for ultracold scattering 
around a magnetic Feshbach
resonance. It includes a relationship between the (generalized)
effective range and the (generalized) scattering length that is
applicable to both broad and narrow resonances, and resonances
of arbitrary $l$. It substantially extends a previous 
relationship \cite{gao98b,fla99,fu03b,gao09a}
that is applicable only to broad resonances.
Two special cases of interest in cold-atom physics,
the case of infinite scattering length (the unitarity limit) and the case
of zero scattering length, are also discussed in this section as examples
of the QDT expansion.
The conclusions are given in Sec.~\ref{sec:conclude}.

\section{Reduction of a multichannel problem to an effective single-channel problem}
\label{sec:reduce}

In cold-atom physics, most of the interest in atomic interaction lies in a small 
range of energies around the lowest threshold (of a certain symmetry), 
below which we have true bound states. 
Ignoring weak couplings between different partial waves due to the magnetic
dipole-dipole \cite{sto88,moe95} and the
second-order spin-orbit interaction \cite{mie96,kot00,leo00},
we can label the single channel of partial wave $l$ that is associated with
this lowest threshold ``$a$'', and all the other channels of partial wave $l$ by ``$c$''.
Above the lowest threshold and below the energies at which the second or 
more channels becomes open, it is already clear from Ref.~\cite{gao05a} that
the MQDT for atom-atom interaction reduces to an effective single-channel
problem with an effective short-range K-matrix, $K^c$, given by
\begin{equation}
K^c_{\text{eff}} = K^{c}_{aa}+K^{c}_{ac}(\chi^c_{cc} - K^{c}_{cc})^{-1}K^{c}_{ca} \;.
\label{eq:Kceff}
\end{equation}
Here $K^{c}_{aa}$, $K^{c}_{ac}$, $K^{c}_{ca}$, $K^{c}_{cc}$,
are submatrices of $K^{c}$ corresponding to 
the separation of all channels into a single ``$a$'' channel and 
$N_c$ closed ``$c$'' channels.
$\chi^c_{cc}$ is an $N_c\times N_c$ diagonal 
matrix with elements $\chi^{c}_l(\epsilon_{si})$,
which is the universal $\chi^{c}_l$ function,
as given, e.g., by Eq.~(54) in paper I, evaluated
at properly scaled channel energies. 
This reduction to an effective single-channel problem
is a result of the standard channel-closing procedure, and
occurs in similar fashion in any type of multichannel
scattering theories. 
What is important, and maybe less well-known, is that the energies of 
the multichannel bound states below the threshold ``$a$''
can also be reduced to an effective single-channel problem
with the \textit{very same} effective $K^c$ as given by Eq.~(\ref{eq:Kceff}). 
A proof is given in the Appendix~\ref{sec:reduceDetail}.

With this reduction, the scattering below the second threshold
and the multichannel bound states below the threshold ``$a$''
are all described by an effective single-channel 
QDT \cite{gao98b,gao01,gao08a} with an effective short-range
$K^c$ matrix given by $K^{c}_{\text{eff}}$. 
Specifically
\begin{equation}
K_l = \tan\delta_l = ( Z^{c}_{gc}K^{c}_{\text{eff}}-Z^{c}_{fc} )
	(Z^{c}_{fs} - Z^{c}_{gs}K^{c}_{\text{eff}})^{-1} \;,
\label{eq:Kphy}	
\end{equation}
gives the scattering $K$ matrix between the lowest and the
second thresholds, and the solutions of (see Appendix~\ref{sec:reduceDetail})
\begin{equation}
\chi^{c}_l(\epsilon_s) = K^{c}_{\text{eff}} \;,
\label{eq:bsp}
\end{equation}
give the bound spectrum below the lowest threshold.
Here $Z^{c}_{xy}(\epsilon_s,l)$ are universal QDT functions
for $-1/r^6$ potential, as given, e.g., by Eqs.~(4)-(7) of
paper I. They, and $\chi^{c}_l(\epsilon_s)$, are all evaluated
at a scaled energy relative to the lowest threshold of 
angular momentum $l$, $\epsilon_s=\epsilon/s_E=(E-E_a)/s_E$, with 
$s_E = (\hbar^2/2\mu)(1/\beta_6)^2$ being the energy scale, 
and $\beta_6=(2\mu C_6/\hbar^2)^{1/4}$ being the length scale 
associated with the $-C_6/r^6$ van der Waals interaction in channel ``$a$''.
We note that other than the ignorance of weak interactions that couple
different $l$ states, there is no further approximation
associated with this reduction to a single channel.

This effective single-channel problem differs from a true
single-channel problem in that the energy dependence of
$K^{c}_{\text{eff}}$ is generally not negligible,
unlike the $K^c$ parameter for a single channel \cite{gao01}. 
As will become clear throughout this work, it is
this energy dependence, which originates from the energy dependence
of $\chi^c_{cc}$ in Eq.~(\ref{eq:Kceff}), that leads
to deviations from single-channel universal behaviors
of paper I \cite{gao09a}, and makes the behaviors of
a ``narrow'' Feshbach resonance to differ substantially from
those of a ``broad'' Feshbach resonance.
As another difference from a true single-channel problem,
the $l$ dependence of $\chi^c_{cc}$ also makes $K^{c}_{\text{eff}}$
$l$-dependent.
The same formalism applies to atomic interaction in an
external magnetic field \cite{han09}, which has the additional effect of
making $K^{c}_{\text{eff}}$ to depend parametrically on $B$.
We will use the notation of $K^{c}_{\text{eff}}(\epsilon,l,B)$,
when necessary, to fully specify its dependences.

The equivalence, around the lowest threshold, 
of the multichannel atomic interaction in a $B$ field to
an effective single channel problem with an effective short-range
$K^{c}_{\text{eff}}(\epsilon,l,B)$ makes most results
of paper I \cite{gao09a} immediately applicable, except for
a few that made explicit use of
the energy- and/or $l$-insensitivity of $K^c$.
In particular, if we define a $K^{c0}_l$ parameter, 
which is more convenient for descriptions
of near-threshold properties \cite{gao08a,gao09a}, as
\begin{equation}
K^{c0}_l(\epsilon,B) = \frac{K^c_{\text{eff}}(\epsilon, l,B)-\tan(\pi\nu_0/2)}
	{1+\tan(\pi\nu_0/2)K^c_{\text{eff}}(\epsilon, l,B)} \;,
\label{eq:Kc0eff}
\end{equation}
where $\nu_0 = (2l+1)/4$ for $-1/r^6$ type of potential,
the locations of the zero-energy
magnetic Feshbach resonances in an arbitrary partial wave $l$,
$B_{0l}$, namely the magnetic fields corresponding to having a 
bound or quasibound state right at the threshold,
can be conveniently found as the roots of 
$K^{c0}_l(\epsilon=0,B)$ \cite{gao00,gao09a}, 
namely as the solutions of
\begin{equation}
K^{c0}_l(\epsilon=0,B_{0l})=0 \;.
\label{eq:B0l}
\end{equation}
The scattering length or 
the generalized scattering length for an abitrary $l$
is given by the zero-energy value of $K^{c0}_l$ through
Eq.~(48) of paper I \cite{gao09a}, namely,
\begin{equation}
\widetilde{a}_l(B) = \bar{a}_{l}\left((-1)^l+\frac{1}{K^{c0}_l(\epsilon=0,B)}\right) \;,
\label{eq:galB}
\end{equation}
where $\bar{a}_{l}=\bar{a}_{sl}\beta_6^{2l+1}$ is the mean scattering
length (with scale included) for partial wave $l$ that was 
defined in paper I \cite{gao09a}, with
\begin{equation}
\bar{a}_{sl} = \frac{\pi^2}{2^{4l+1}[\Gamma(l/2+1/4)\Gamma(l+3/2)]^2} \;,
\end{equation}
being the scaled mean scattering length. 
Recall that the generalized scattering length, $\widetilde{a}_l$,
reduces to the regular scattering lengths whenever they are well
defined, namely for the $s$ and $p$ partial waves.

Computationally, a similar theory based on MQDT \cite{gao05a}
has been shown by Hanna \textit{et al.} \cite{han09} to give an accurate description
of magnetic Feshbach resonances over a wide range of magnetic
fields using only three parameters for alkali-metal systems.
Even better results can be expected by incorporating the energy
and/or partial-wave dependences of $K^c_S$ and $K^c_T$ \cite{gao05a} 
using a few more parameters.
Further calculations for specific systems and especially nonzero partial
waves will be presented elsewhere.
Here we focus on the parametrization of one particular resonance
and the analytic description of atomic interaction
around it.

\section{Parametrization of a magnetic Feshbach resonance}
\label{sec:para}

\subsection{Derivation and general considerations}

As the second step towards developing an analytic description 
of a magnetic Feshbach resonance, we need a simple parametrization
of $K^c_{\text{eff}}$ or the corresponding $K^{c0}_l$. 
For any isolated resonance, the second term in Eq.~(\ref{eq:Kceff})
has a simple pole at $\bar{\epsilon}_l(B)$, determined by 
$\det(\chi^c - K^{c}_{cc})=0$. It represents the ``bare'' location 
of a Feshbach resonance and depends on the magnetic field.
(Here ``bare'' means no coupling to the open channel ``a''.)
Around such a simple pole, 
the effective $K^c$ parameter, Eq.~(\ref{eq:Kceff}), 
can always be parametrized as
\begin{equation}
K^c_{\text{eff}} = K^c_{\text{bg}l}-\frac{\Gamma^c_{l}/2}{\epsilon-\bar{\epsilon}_l(B)} \;,
\label{eq:Kcpa}
\end{equation}
sufficiently close to the pole.
Here $\Gamma^c_{l}$ is a measure of the width of the resonance, 
$\epsilon$ and $\bar{\epsilon}_l$ are energies that are conveniently chosen
to be relative to channel ``$a$'', e.g., $\epsilon=E-E_a$,
and $K^c_{\text{bg}l}$ is a background $K^c$ parameter, namely the $K^c$ for
energies and magnetic fields away from the resonance,
such that $|\epsilon-\bar{\epsilon}_l(B)|\gg \Gamma^c_{l}$.
Using the fact that $\chi^c_l$ is a piecewise monotonically decreasing functions
of energy \cite{gao01}, namely, $d\chi^c_l/d\epsilon_s<0$, one can further show rigorously
that $\Gamma^c_{l}>0$, a property that will put important constraints
on other forms of parametrizations, all of which will be derived from Eq.~(\ref{eq:Kcpa}).

In writing Eq.~(\ref{eq:Kcpa}), we have adopted a notation that avoids
unnecessary confusions without getting into the details of the MQDT \cite{gao05a}
for atomic interaction in a magnetic field \cite{han09}.
Rigorously speaking, the $K^c$ matrix itself, and therefore the parameters
$K^c_{\text{bg}l}$ and $\Gamma^c_{l}$ in Eq.~(\ref{eq:Kcpa}), also depend on $B$.
This dependence, however, is only significant over a field range of
the order of $\Delta E^{\text{hf}}/\mu_B$, where $\Delta E^{\text{hf}}$ 
is the atomic hyperfine
splitting, and $\mu_B$ is the Bohr magneton.
Since our focus here is on the parametrization of an individual resonance, the
width of which is always much smaller than the hyperfine splitting \cite{chi10},
we adopt the notation of Eq.~(\ref{eq:Kcpa}) to emphasize that over
the range of $B$ field of interest here, the most relevant
$B$ field dependence is that of the ``bare'' Feshbach energy, $\bar{\epsilon}_l(B)$.

As discussed in paper I \cite{gao09a}, analytic properties around the threshold are
more conveniently described using the short-range parameter $K^{c0}_l$.
Substituting Eq.~(\ref{eq:Kcpa}) into Eq.~(\ref{eq:Kc0eff}),
we have
\begin{equation}
K^{c0}_{l}(\epsilon,B) = K^{c0}_{\text{bg}l}
	-\frac{\Gamma^{c0}_l/2}
	{\epsilon-\bar{\epsilon}_l(B)-f_{El}}\;,
\label{eq:Kc0pag}
\end{equation}
where $K^{c0}_{\text{bg}l}$ is the background $K^{c0}_l$ parameter 
corresponding to $K^c_{\text{bg}l}$
\begin{equation}
K^{c0}_{\text{bg}l} = \frac{K^c_{\text{bg}l}-\tan(\pi\nu_0/2)}{1+\tan(\pi\nu_0/2)K^c_{\text{bg}l}}\;,
\label{eq:Kc0bgl}
\end{equation}
with a corresponding generalized background scattering length of \cite{gao09a}
\begin{equation}
\widetilde{a}_{\text{bg}l} = \bar{a}_{l}\left((-1)^l+\frac{1}{K^{c0}_{\text{bg}l}}\right) \;,
\label{eq:gabgl}
\end{equation}
and
\begin{equation}
\Gamma^{c0}_l = \Gamma^c_l\frac{1+\tan^2(\pi\nu_0/2)}
	{[1+\tan(\pi\nu_0/2)K^c_{\text{bg}l}]^2} \;.
\label{eq:Gamc0l}	
\end{equation}
The $f_{El}$ in Eq.~(\ref{eq:Kc0pag}) is not an independent parameters. It is related
to $K^{c0}_{\text{bg}l}$ and $\Gamma^{c0}_l$ by
\begin{equation}
f_{El} = \frac{1}{2}\Gamma^{c0}_l\frac{\tan(\pi\nu_0/2)}
	{1-\tan(\pi\nu_0/2)K^{c0}_{\text{bg}l}} \;.
\label{eq:fEl}	
\end{equation}
In describing a Feshbach resonance in terms of $K^{c0}_l$, 
the fact that $\Gamma^{c}_l>0$ translates into the condition of 
$\Gamma^{c0}_l>0$, as is clear from Eq.~(\ref{eq:Gamc0l}).

From Eq.~(\ref{eq:Kc0pag}), the most general parametrization of 
a magnetic Feshbach resonance is that of Appendix~\ref{sec:altpara}.
We adopt here a slightly less general parametrization
that uses parameters that have more direct physical
interpretations, and are more closely aligned with those
already well established for the $s$ wave \cite{moe95,koh06,chi10}.

For $K^{c0}_{\text{bg}l}\neq 0$ ($\widetilde{a}_{\text{bg}l}\neq \infty$), 
namely in all cases when there is no 
background bound or quasi-bound state right at the threshold \cite{gao00},
it is more convenient to rewrite Eq.~(\ref{eq:Kc0pag}) as
\begin{eqnarray}
K^{c0}_{l}(\epsilon,B) &=& K^{c0}_{\text{bg}l}\frac{\epsilon-\epsilon_l(B)}
	{\epsilon-\epsilon_l(B)-d_{El}}\;, \\
	&=& K^{c0}_{\text{bg}l}\left(1+\frac{d_{El}}
	{\epsilon-\epsilon_l(B)-d_{El}}\right) \;,
\label{eq:Kc0l1}
\end{eqnarray}
where
\begin{equation}
\epsilon_l(B) = \bar{\epsilon}_l(B)+\frac{\Gamma^c_{l}/2}{K^c_{\text{bg}l}-\tan(\pi\nu_0/2)} \;,
\label{eq:elB}
\end{equation}
and 
\begin{equation}
d_{El} = (\Gamma^c_{l}/2)\frac{1+\tan^2(\pi\nu_0/2)}{[\tan(\pi\nu_0/2)-K^c_{\text{bg}l}]
	[1+\tan(\pi\nu_0/2)K^c_{\text{bg}l}]} \;.
\end{equation}
In this form for $K^{c0}_l$, the location of the zero-energy magnetic 
Feshbach resonance, $B_{0l}$, determined by Eq.~(\ref{eq:B0l}),
translates into the solution of $\epsilon_l(B_{0l})=0$.
And since we are interested here only in a range of $B$ that
covers a single Feshbach resonance,
the $\epsilon_l(B)$ in Eq.~(\ref{eq:Kc0l1}) can be approximated, 
around $B_{0l}$, by
$\epsilon_l(B)\approx \delta\mu_l(B-B_{0l})$,
where $\delta\mu_l = \left.d\epsilon_l(B)/dB\right|_{B=B_{0l}}$ 
is the difference 
of magnetic moments between the molecular state and the 
separate-atom state \cite{chi10}.
This approximation, together with Eq.~(\ref{eq:Kc0l1}), 
gives the following parameterization of the
effective $K^{c0}_l$ around a magnetic Feshbach resonance,
\begin{equation}
K^{c0}_{l}(\epsilon,B) = 
	K^{c0}_{\text{bg}l}\left(1+\frac{d_{El}}
	{\epsilon-\delta\mu_l(B-B_{0l})-d_{El}} \right) \;.
\label{eq:Kc0Fesh}	
\end{equation}
It is a parametrization 
in terms of four parameters $B_{0l}$, $K^{c0}_{\text{bg}l}$, $\delta\mu_l$, and $d_{El}$,
with the condition of $K^{c0}_{\text{bg}l}d_{El}<0$ due to $\Gamma^{c0}_l>0$.
These parameters, together with either the $C_6$ coefficient or the corresponding
energy scale $s_E$ for a total of five parameters, 
provide a complete characterization of atomic interaction 
around a magnetic Feshbach resonance, through Eqs.~(\ref{eq:Kphy}) and (\ref{eq:bsp}).
It is applicable for all partial waves $l$ and for either broad or 
narrow Feshbach resonances (the 
precise definition of which will be addressed in in Sec.~\ref{sec:scaledpara}), 
or anything in between.
It fails only in the special case of having a \textit{background}
bound or quasibound state right at the threshold, which
can happen only by pure coincidence.
This special case, together with an alternative parametrization
of magnetic Feshbach resonances that is applicable for all cases,
is discussed in Appendix~\ref{sec:altpara}.

\subsection{Tuning of the scattering lengths and generalized scattering lengths}
\label{sec:galB}

Contained in the parametrization of $K^{c0}_l$ is the
magnetic-field dependence of 
the scattering length or 
the generalized scattering length for an abitrary $l$.
Defining $K^{c0}_l(B)\equiv K^{c0}_{l}(\epsilon=0,B)$ to simplify
the notation, we have from Eq.~(\ref{eq:Kc0Fesh})
\begin{equation}
K^{c0}_{l}(B) = 
	K^{c0}_{\text{bg}l}\left(1-\frac{d_{Bl}}{B-B_{0l}+d_{Bl}}\right) \;,
\label{eq:Kc0Fesh0}	
\end{equation}
where $d_{Bl}=d_{El}/\delta\mu_l$.
Upon substitution into Eq.~(\ref{eq:galB}), we obtain,
for $\widetilde{a}_{\text{bg}l}\neq 0$, 
\begin{eqnarray}
\widetilde{a}_{l}(B) &=& \bar{a}_{l}\left((-1)^l+\frac{1}{K^{c0}_{l}(B)}\right) \label{eq:gaBold1}\\
	&=& \widetilde{a}_{\text{bg}l}\left(1-\frac{\Delta_{Bl}}{B-B_{0l}}\right) \;.
\label{eq:gaBold}
\end{eqnarray}
Here $\widetilde{a}_{\text{bg}l}$ is the (generalized) background scattering length
defined earlier by Eq.~(\ref{eq:gabgl}), and
\begin{eqnarray}
\Delta_{Bl} &=& -d_{Bl}/[1+(-1)^lK^{c0}_{bgl}] \;,\\
	&=& -\left(1-(-1)^l\frac{1}{\widetilde{a}_{\text{bg}l}/\bar{a}_l}\right)d_{Bl} 
	\label{eq:DBl}\;.
\end{eqnarray}
For $\widetilde{a}_{\text{bg}l}=0$, we obtain
\begin{equation}
\widetilde{a}_{l}(B) = -(-1)^{l}\bar{a}_l\frac{d_{Bl}}{B-B_{0l}} \;.
\label{eq:gaBbg0}
\end{equation}
Equation~(\ref{eq:gaBold}) shows that around a magnetic Feshbach resonance
of arbitrary $l$, the (generalized) scattering length is tuned in a similar
fashion by the magnetic field as around an $s$ wave resonance,
and can be parametrized in a similar manner \cite{moe95,koh06,chi10}.

The parametrization of the $s$ wave scattering length in the form of
Eq.~(\ref{eq:gaBold}) has been popular for a good reason: every parameter in it
has the simplest and the most direct experimental interpretation.
It is worth pointing out, however, that theoretically it is not the most general
parametrization possible as it fails for both 
$\widetilde{a}_{\text{bg}l}=\infty$ and $\widetilde{a}_{\text{bg}l}=0$.
As discussed in Appendix~\ref{sec:altpara},
the failure of Eq.~(\ref{eq:gaBold}) at $\widetilde{a}_{\text{bg}l}=\infty$,
and the corresponding failure of Eqs.~(\ref{eq:Kc0Fesh}) and 
(\ref{eq:Kc0Fesh0}) at $K^{c0}_{\text{bg}l}=0$,
is a necessary sacrifice for using $B_{0l}$, which has a more direct
physical interpretation than the $\bar{B}_{0l}$ parameter 
of  Appendix~\ref{sec:altpara}, 
but does not exist for $\widetilde{a}_{\text{bg}l}=\infty$.
Its failure at $\widetilde{a}_{\text{bg}l}=0$ is the price we
pay for using the parameter $\Delta_{Bl}$.
An alternative parametrization of the scattering length,
which remains applicable for $\widetilde{a}_{\text{bg}l}=0$, is
\begin{equation}
\widetilde{a}_{l}(B) = \widetilde{a}_{\text{bg}l}
	+\frac{\widetilde{a}_{\text{bg}l}-(-1)^{l}\bar{a}_l}{(B-B_{0l})/d_{Bl}} \;.
\label{eq:gaB}	
\end{equation}
It can be obtained, e.g., by substituting Eq.~(\ref{eq:DBl}) for 
$\Delta_{Bl}$ into Eq.~(\ref{eq:gaBold}).
This parametrization is well defined, and reduces to
Eq.~(\ref{eq:gaBbg0}) for $\widetilde{a}_{\text{bg}l}=0$.

\subsection{Parametrization in terms of scaled parameters
and the definitions of ``broad'' and ``narrow'' resonances}
\label{sec:scaledpara}

Of the five parameters required to completely characterize
the atomic interaction around a Feshbach resonance,
such as $B_{0l}$, $K^{c0}_{\text{bg}l}$, $d_{El}$, $\delta\mu_l$,
and $s_E$ (or $C_6$), two of them, $K^{c0}_{\text{bg}l}$ and $d_{El}$
can be replaced by $\widetilde{a}_{\text{bg}l}$ and $\Delta_{Bl}$,
used in the parametrization of the
(generalized) scattering length.
The resulting parametrization, in terms of 
$B_{0l}$, $\widetilde{a}_{\text{bg}l}$, $\Delta_{Bl}$, 
$\delta\mu_l$ and $s_E$ (or $C_6$), gives an alternative that is 
the most direct generalization 
of the $s$ wave parametrization \cite{chi10} 
to other partial waves.
Both sets, however, have the limitation that
they are not fully transparent to the distinction between broad
and narrow resonances.

The effective single channel $K^{c0}_l$ parameter for
a magnetic Feshbach resonance, as characterized, e.g., by
Eq.~(\ref{eq:Kc0Fesh}), is generally energy-dependent.
Depending on the relative importance of this energy variation, 
as compared to those due to the long-range van der Waals 
interaction, a Feshbach resonance can be classified either
as ``broad'' or ``narrow''. For a broad Feshbach resonance,
the energy dependence of $K^{c0}_l$ is insignificant compared
to those induced by the van der Waals interaction. The atomic
interaction around such a resonance follows, to a large extent,
the single-channel universal behavior of paper I \cite{gao09a}
with a tunable (generalized) scattering length.
A narrow Feshbach resonance corresponds to the opposite limit
in which the energy dependence of $K^{c0}_l$ dominates.
The atomic interaction around such a resonance can differ 
completely from the single-channel universal behavior.

\begin{figure}
\scalebox{0.4}{\includegraphics{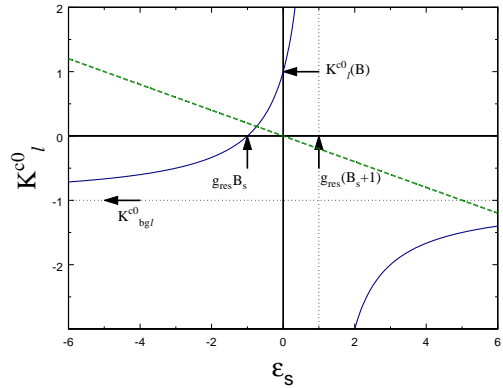}}
\caption{(Color online) Illustrations of the parameters describing 
the energy dependence of $K^{c0}_l(\epsilon_s,B_s)$
on the van der Waals energy scale. 
$\Gamma^{c0}_l>0$ ($K^{c0}_{\text{bg}l}g_{\text{res}}<0$) implies that
$K^{c0}_l$ is piecewise monotonically increasing function of energy.
The significance of its energy dependence is determined
by comparing it with that induced by the van der Waals interaction,
the order-of-magnitude of which can be measured by the
energy dependence of the $\theta_l\approx -\epsilon_s/(2l+3)(2l-1)$
function. The dashed line illustrates the $\theta_l$ function
for $l=1$. The energy variation of $\theta_l$ is less significant for
higher partial waves.
\label{fig:Kc0}}
\end{figure}

To better characterize the relative importance of the
energy dependence of $K^{c0}_l$ and therefore the definition
of broad and narrow resonances, we need to first
put it on the same energy scale as the
other energy-dependent functions, namely on the energy scale
$s_E = (\hbar^2/2\mu)(1/\beta_6)^2$ that is associated with
the van der Waals interaction \cite{gao09a}.
Defining
\begin{equation}
g_{\text{res}} = d_{El}/s_E \;,
\end{equation}
and
\begin{equation}
B_s = (B-B_{0l})/d_{Bl} \;,
\end{equation}
Eq.~(\ref{eq:Kc0Fesh}) can be written as
\begin{equation}
K^{c0}_l(\epsilon_s,B_s) = K^{c0}_{\text{bg}l}
	\left(1+\frac{g_{\text{res}}}{\epsilon_s-g_{\text{res}}(B_s+1)}\right) \;.\\
\label{eq:Kc0Feshs1}	
\end{equation}	
It describes $K^{c0}_l$ as a function of the scaled energy $\epsilon_s$ and
a scaled magnetic field $B_s$ using two dimensionless parameters,
$K^{c0}_{\text{bg}l}$ and $g_{\text{res}}$, the meaning of which are 
illustrated in Fig.~\ref{fig:Kc0}.
$K^{c0}_{\text{bg}l}$ is the background $K^{c0}_l$, namely its value away
from the resonance. $g_{\text{res}}$ is a measure of the width
of the resonance. More specifically,
$K^{c0}_l(\epsilon_s,B)$ goes to infinity at $\epsilon_s=g_{\text{res}}(B_s+1)$.
It crosses zero $\epsilon_s=g_{\text{res}}B_s$. The distance between
the two locations is $|g_{\text{res}}|$, which measures the width of resonance.
The parametrization of $K^{c0}_l$ using Eq.~(\ref{eq:Kc0Feshs1})
divide the parameters characterizing a magnetic Feshbach resonance
into three parameters, $B_{0l}$, $d_{Bl}$, and $s_E$, for location and scaling, 
and two dimensionless parameters, $K^{c0}_{\text{bg}l}$ and $g_{\text{res}}$, for the shape.
Feshbach resonances with the same shape parameters differ from
each other only in scaling.
The condition of $\Gamma^{c0}_l>0$ implies that
the two shape parameters $K^{c0}_{\text{bg}l}$ and $g_{\text{res}}$ 
are constrained by $K^{c0}_{\text{bg}l}g_{\text{res}}<0$.

With the parametrization given by Eq.~(\ref{eq:Kc0Feshs1}), 
we are now in position for rigorous definitions of broad
and narrow resonances. The $g_{\text{res}}$ parameter,
which measures the width of the resonance on the scale of $s_E$,
gives a rough, yet still imprecise, classification of ``broad'' 
($|g_{\text{res}}|\gg 1$) and ``narrow'' ($|g_{\text{res}}|\ll 1$).
It is not precise because the energy variation due to the van der 
Waals interaction over a scale of $s_E$ is different for different
partial waves. This is especially true for large $l$, for which
the energy variation around the threshold due to the van der Waals 
interaction is much less significant than that for the $s$ wave.

For a more precise definition of ``broad'' and ``narrow'',
we first recognize that the leading energy
variation due to the van der Waals interaction is characterized
by the $\theta_l\approx -\epsilon_s/(2l+3)(2l-1)$ function defined
by the Eq.~(35) of paper I \cite{gao09a} (repeated as 
Eq.~(\ref{eq:thetal}) in Sec.~\ref{sec:scatteringB}).
This energy variation, as measured by 
$|\partial \theta_l/\partial\epsilon_s(\epsilon_s=0)| = |1/(2l+3)(2l-1)|$,
is what should be compared with the energy variation of the $K^{c0}_l$
at zero energy, as measured by 
$\partial K^{c0}_l/\partial\epsilon_s(\epsilon_s=0,B_s=0) = -K^{c0}_{\text{bg}l}/g_{\text{res}}$.
This leads to the definition of an auxiliary parameter
\begin{equation}
\zeta_{\text{res}}\equiv \frac{g_{\text{res}}}{(2l+3)(2l-1)K^{c0}_{\text{bg}l}} \;,
\end{equation}
which gives a precise characterization of ``broad'' and ``narrow''.
For a broad resonance with $|\zeta_{\text{res}}|\gg 1$, the energy
variation of the effective short-range parameter is insignificant
compared to that due to the van der Waals interaction, just like
the case of a single channel \cite{gao09a}. The atomic interaction
around such a resonance can be expected
to follow the single channel universal behavior.
For a narrow resonance with $|\zeta_{\text{res}}|\ll 1$, the energy
variation of the effective short-range parameter dominates.
Within such a resonance, the atomic interaction deviates substantially
from the single channel behavior. The constraint $K^{c0}_{\text{bg}l}g_{\text{res}}<0$
implies that $\zeta_{\text{res}}$
is always positive for $l=0$, and always negative for $l\ge 1$.
Specializing to the $s$ wave, the $\zeta_{\text{res}}$ parameter
is similar in spirit to the parameter $s_{\text{res}}$ of 
Chin \textit{et al.} \cite{chi10}, which is equivalent to the $1/\eta$ 
parameter of K\"{o}hler \textit{et al.} \cite{koh06}.

\begin{table*}
\caption{Sample parameters for selective $s$ wave magnetic Feshbach resonances,
illustrating a vast range of $\zeta_{\text{res}}$ values, from very narrow
($|\zeta_{\text{res}}|\ll 1$) to very broad ($|\zeta_{\text{res}}|\gg 1$).
It shows, for example, that $^6$Li-$^6$Li and $^{133}$Cs-$^{133}$Cs systems
have the best resonances for the purpose of investigating universal behaviors.
Here $a_0$ is the Bohr radius and $\mu_B$ is the Bohr magneton. The data sets of
$B_{0l}$, $\Delta_{Bl}$, $a_{\text{bg}l}$, and $\delta\mu_l$ are taken from
Table~IV of Chin \textit{et al.} \cite{chi10}. The channel identification
also follows the same reference. Note that there are many resonances that
are not broad. The atomic interaction around them do not follow single-channel
universal behavior and is much better described using the QDT expansion 
presented here.}
\begin{ruledtabular}
\begin{tabular}{lrcr|rrr|rrr|r|l}
  system       & $s_E/k_B$($\mu$K) & ch.  & $B_{0l}$(G) & $\Delta_{Bl}$(G) & $a_{\text{bg}l}/a_0 $ & $\delta\mu_l/\mu_{B}$ & $K^{c0}_{\mathrm{bg}l}$ & $g_\mathrm{res}$ & $d_{Bl}$(G) & $\zeta_{\text{res}}$ & references\\ \hline
  $^6$Li$^6$Li & 7368        & $ab$ &  834.1 & -300    & -1405 & 2.0   & -0.02083 & 5.356 & 293.8  & 85.73 & \cite{bar05}\\
               &              & $ac$ &  690.4 & -122.3  & -1727 & 2.0   & -0.01701 & 2.192 & 120.2  & 42.96 & \cite{bar05}\\
               &              & $bc$ &  811.2 & -222.3  & -1490 & 2.0   & -0.01966 & 3.974 & 217.9  & 67.37 & \cite{bar05}\\
               &              & $ab$ &  543.25 & 0.1     & 60    & 2.0   & 0.9923   & -0.00363 & -0.1992 & 0.00122 & \cite{str03}  \\

  $^7$Li$^7$Li & 5849        & $aa$ &  736.8  & -192.3      & -25   & 1.93   & -0.5540  & 1.901 & 85.76  & 1.144 & \cite{pol09,str02,chi10} \\    
  \hline
  $^{23}$Na$^{23}$Na & 933.1  & $cc$ &  1195    & -1.4      & 62    & -0.15   & 2.255    & -0.04921 & 4.557 & 0.00727  & \cite{ino98,ste99,chi10} \\
                     &        & $aa$ &  907    & 1       & 63    & 3.8   & 2.143    & -0.8597 & -3.143 & 0.1337  & \cite{ino98,ste99,chi10} \\
                     &        & $aa$ &  853    & 0.0025  & 63    & 3.8   & 2.143    & -0.00215 & -0.00786 & 0.00033 & \cite{ino98,ste99,chi10} \\    \hline

  $^{40}$K$^{40}$K   & 257.3  & $ab$ &  202.1  & 8.0     & 174   & 1.68  & 0.5542 & -5.454 & -12.43 & 3.280 & \cite{reg04,chi10} \\
                     &        & $ac$ &  224.2  & 9.7     & 174   & 1.68  & 0.5542 & -6.613 & -15.07 & 3.978 & \cite{reg03b,chi10} \\    \hline

  $^{85}$Rb$^{85}$Rb & 75.58  & $ee$ &  155.04 & 10.7   & -443   & -2.33 & -0.1506  & 18.82 & -9.089 & 41.66 & \cite{cla03} \\

  $^{87}$Rb$^{87}$Rb & 72.99  & $aa$ &  1007.4 & 0.21   & 100   & 2.79 & 3.759   & -2.566 & -0.9994  & 0.2275 & \cite{vol03,dur04,chi10} \\
               & &$aa$ &  911.7  & 0.0013 & 100   & 2.71 & 3.759  & -0.01543  & -0.00619 & 0.00137 & \cite{mar02,chi10} \\
               & &$aa$ &  685.4  & 0.006  & 100   & 1.34 & 3.759  & -0.03521  & -0.02855 & 0.00312 & \cite{mar02,dur04,chi10}\\
               & &$aa$ &  406.2  & 0.0004 & 100   & 2.01 & 3.759  & -0.003521 & -0.00190 & 0.00031  & \cite{mar02,chi10} \\
               & &$ae$ &  9.13   & 0.015  & 99.8  & 2.00 & 3.795  & -0.1324 & -0.07193  & 0.01163 & \cite{wid04} \\        \hline

  $^{133}$Cs$^{133}$Cs & 31.97 & $aa$ &  -11.7 & 28.7  & 1720  & 2.30  & 0.05945 & -146.9 & -30.41  & 823.9 & \cite{chi04,lan09,chi10} \\
               & &$aa$ &  547    & 7.5    & 2500  & 1.79 & 0.04015 & -29.34 & -7.801  & 243.5 & \cite{chi10} \\
               & &$aa$ &  800    & 87.5   & 1940  & 1.75 & 0.05235 & -338.6 & -92.08  & 2156  & \cite{chi10} \\
\end{tabular}
\end{ruledtabular}
\label{tab:resonances}
\end{table*}
To finish our discussion on parametrization, we summarize here the explicit relations
between two sets of parameters that we will use for the complete
characterization of a Feshbach resonance.
The first set is 
$B_{0l}$, $\widetilde{a}_{\text{bg}l}$, $\Delta_{Bl}$, 
$\delta\mu_l$ and $s_E$ (or $C_6$), which is more closely correlated
with the parametrization of the (generalized) scattering length 
and the standard $s$ wave parametrization \cite{chi10}.
The second set is $B_{0l}$, $K^{c0}_{\text{bg}l}$, $g_{\text{res}}$, 
$d_{Bl}$, and $s_E$ (or $C_6$),
which is much more convenient with the QDT expansion of Sec.~\ref{sec:scatteringB},
and correlates much more closely with the distinction of broad and narrow
resonances. They differ in three parameters that are related by
\begin{equation}
K^{c0}_{\text{bg}l} = \frac{1}{\widetilde{a}_{\text{bg}l}/\bar{a}_{l} -(-1)^l} \;,
\label{eq:Kc0bgga}
\end{equation}
\begin{equation}
g_{\text{res}} = -\frac{\widetilde{a}_{\text{bg}l}/\bar{a}_{l}}{\widetilde{a}_{\text{bg}l}/\bar{a}_{l} -(-1)^l} 
	\left(\frac{\delta\mu_l \Delta_{Bl}}{s_E}\right)\;,
\label{eq:gres}
\end{equation}
\begin{equation}
d_{Bl} = -\frac{\widetilde{a}_{\text{bg}l}/\bar{a}_{l}}{\widetilde{a}_{\text{bg}l}/\bar{a}_{l} -(-1)^l} 
	\Delta_{Bl} \;.
\label{eq:dBl}
\end{equation}
The condition of $\Gamma^c_l>0$ translates into the constraint
$\delta\mu_l\widetilde{a}_{\text{bg}l}\Delta_{Bl}>0$
for the first set of parameters, and into
$K^{c0}_{\text{bg}l}g_{\text{res}}<0$ for the second.
Table~1 gives examples of both sets of parameters for selective $s$ 
wave magnetic Feshbach resonances. 
The first set is taken from Table~IV of Chin \textit{et al.} \cite{chi10}.
The second set is calculated from the first using Eqs.~(\ref{eq:Kc0bgga})-(\ref{eq:dBl}).
They are given here both for convenient applications of the QDT expansion,
and to illustrate the vast range of $\zeta_{\text{res}}$, from very narrow
$|\zeta_{\text{res}}|\ll 1$ to very broad $|\zeta_{\text{res}}|\gg 1$.
Since the parametrization is new for nonzero partial waves,
no parameters are yet available for them.
Tentative theoretical predictions of resonances and their 
parameters for nonzero partial waves,
using MQDT as briefly outlined in Sec.~{\ref{sec:reduce}, 
will be presented elsewhere. 
It is hoped that they will stimulate further experiment and theory 
for their more precise characterization.
Previous works on nonzero partial waves, such as those in
Ref.~\cite{chi00,reg03b,tic04,zha04,sch05,gae07,kno08,fuc08,han09,lys09,chi10},
can also be re-analyzed to extract the parameters.

The second set of parameters describes $K^{c0}_l$ through
Eq.~(\ref{eq:Kc0Feshs1}). A useful variation, which relates $K^{c0}_l$ 
explicitly to its value at zero energy, is given by
\begin{equation}
K^{c0}_l(\epsilon_s,B_s) = \frac{K^{c0}_l(B_s)-K^{c0}_{\text{bg}l}\eta(B_s)\epsilon_s}
	{1-\eta(B_s)\epsilon_s} \;,
\label{eq:Kc0Feshs2}	
\end{equation}	
where
\begin{equation}
K^{c0}_l(B_s) = K^{c0}_l(\epsilon_s=0,B_s) = K^{c0}_{\text{bg}l}\frac{B_s}{B_s+1}\;,
\end{equation}
is the value of $K^{c0}_l$ at zero energy, given earlier by Eq.~(\ref{eq:Kc0Fesh0}),
expressed in terms of the scaled magnetic field,
and we have defined 
\begin{equation}
\eta(B_s) = \frac{1}{g_{\text{res}}(B_s+1)}\;.
\end{equation}
This representation of $K^{c0}_l$
makes it clear that $K^{c0}_l(\epsilon_s,B_s)\sim K^{c0}_l(B_s)$
in the broad-resonance limit of $|g_{\text{res}}|\rightarrow\infty$.
It also makes it easier, if ever desirable, to represent $K^{c0}_l(\epsilon_s,B_s)$
in term of (generalized) scattering
length and (generalized) background scattering length, using 
Eqs.~(\ref{eq:Kc0bgga})-(\ref{eq:dBl}), and
\begin{equation}
K^{c0}_l(B_s) = \frac{1}{\widetilde{a}_l(B)/\bar{a}_{l}-(-1)^l} \;,
\label{eq:Kc0laB}
\end{equation}
which is a direct consequence of Eq.~(\ref{eq:gaBold1}).

\section{QDT expansion for ultracold scattering around a magnetic Feshbach resonance}
\label{sec:scatteringB}

In deriving the QDT expansion for single-channel ultracold scattering
of paper I \cite{gao09a}, the only quantities expanded are the universal
QDT functions, with no assumptions made about the behavior of the
short-range parameter $K^{c0}_l$, including its energy dependence.
Thus the same expansion is applicable to the effective single-channel
problem that describes the magnetic Feshbach resonance.
Specifically, we have for $\epsilon_s>0$ \cite{gao09a},
\begin{equation}
\tan\delta_l \approx K^{(B)}_l(\epsilon_s)+K^{(D)}_l(\epsilon_s,B_s) \;,
\label{eq:qdtexpB1}	
\end{equation}
where
\begin{eqnarray}
K^{(B)}_l &\approx& -\pi(\nu-\nu_0) \nonumber\\
	&\approx& \frac{3\pi}{(2l+5)(2l+3)(2l+1)(2l-1)(2l-3)}\epsilon_s^{2} \;,
\label{eq:qdtexpB2}	
\end{eqnarray}
is the Born term (see, e.g., Ref.~\cite{lan77}), and
\begin{equation}
K^{(D)}_l(\epsilon_s,B_s) = -\widetilde{A}_{sl}(\epsilon_s,B_s)k_s^{2l+1} \;,
\label{eq:qdtexpB3}	
\end{equation}
describes the deviation from the Born term. Here
\begin{widetext}
\begin{eqnarray}
\widetilde{A}_{sl}(\epsilon_s,B_s) &=& \bar{a}_{sl}\left[
	(-1)^l+\frac{1+K^{c0}_l(\epsilon_s,B_s)\theta_l}
	{K^{c0}_l(\epsilon_s,B_s)-\theta_l-\pi(\nu-\nu_0)/2}\right] \;,
\label{eq:qdtexpB4}	\\
	&=& \bar{a}_{sl}\left[
	(-1)^l+\frac{(2l+3)(2l-1)-K^{c0}_l(\epsilon_s,B_s)\epsilon_s}
	{(2l+3)(2l-1)K^{c0}_l(\epsilon_s,B_s)+\epsilon_s+w_l\epsilon_s^2}\right] \;,
\label{eq:qdtexpB5}	
\end{eqnarray}
\end{widetext}
with the $w_l$ in Eq.~(\ref{eq:qdtexpB5}) being given by
\begin{equation}
w_l = \frac{3\pi}{2(2l+5)(2l+1)(2l-3)} \;,
\end{equation}
and the $\theta_l$ in Eq.~(\ref{eq:qdtexpB4}) being given by
\begin{equation}
\theta_l \approx -\frac{1}{(2l+3)(2l-1)}\epsilon_s \;.
\label{eq:thetal}
\end{equation}

Equations~(\ref{eq:qdtexpB1})-(\ref{eq:qdtexpB5}),
with a $K^{c0}_l$ that depends explicitly on energy and
parametrically on the magnetic field $B$, as described by either 
Eq.~(\ref{eq:Kc0Feshs1}), Eq.~(\ref{eq:Kc0Feshs2}), or 
Eq.~(\ref{eq:Kc0la}) of Appendix~\ref{sec:altpara},
give the QDT expansion for scattering around a magnetic 
Feshbach resonance in an arbitrary partial wave $l$.
It is applicable to both broad
and narrow resonances, or anything in between, and
has the same energy range of applicability as its single-channel
counterpart, limited only
by $\epsilon_s$ being much less than the critical scale energy 
$\epsilon_{scl}$, as discussed in more detail in paper I \cite{gao09a}.
There is no restriction on the magnetic field except that imposed by the
validity of the isolated resonance. 

\begin{figure}
\scalebox{0.4}{\includegraphics{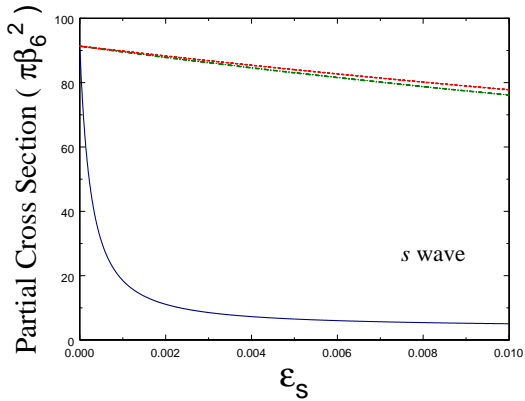}}
\caption{(Color online) Comparison of near-threshold 
$s$ wave scattering properties for narrow and broad
Feshbach resonances. 
The solid line represents results for 
the $^6$Li $ab$ channel narrow resonance located at 543 G.
The dash-dot line represents results for
the $^6$Li $ab$ channel broad resonance located at 834 G.
In both cases, magnetic fields are chosen to give the same 
$s$ wave scattering length corresponding to 
$a_{l=0}(B)/\bar{a}_{l=0}=+10$.
The figure also shows that the Feshbach resonance at 834 G
is sufficient broad that the scattering properties around it
is well approximated by the single-channel universal behavior
(dashed line).
\label{fig:sbn1}}
\end{figure}
\begin{figure}
\scalebox{0.4}{\includegraphics{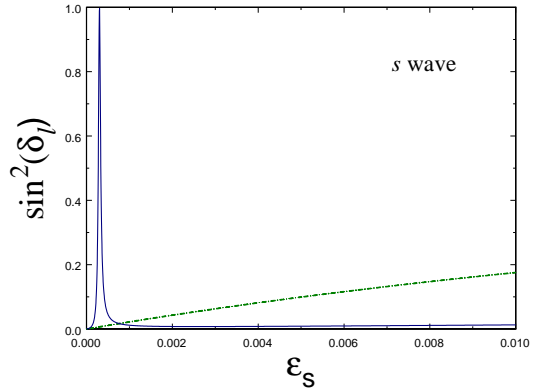}}
\caption{(Color online) The same as Fig.~\ref{fig:sbn1} except it is for 
$a_{l=0}(B)/\bar{a}_{l=0}=-10$.
$\sin^2\delta_l$ is plotted, instead of the partial cross section
to give better visibility to both sets of data on the same figure.
The single-channel universal behavior is indistinguishable from the
broad-resonance results (dash-dot line) and is not plotted.
Note that even though both set of data correspond to the same 
scattering length, the case of narrow Feshbach (solid line) has a 
resonance feature in the threshold region that is absent for
a broad Feshbach. (See also Ref.~\cite{chi10}.)
\label{fig:sbn2}}
\end{figure}
While the QDT expansion for scattering around a magnetic
Feshbach resonance may be formally similar to QDT expansion for true 
single channel cases \cite{gao09a}, 
it contains considerable new physics beyond those
of a single channel, including dramatically different behaviors
for broad and narrow resonances.
For broad resonances with
$|g_{\text{res}}|\gg 1$, or more precisely
$|\zeta_{\text{res}}|\gg 1$, the energy dependence of
$K^{c0}_l(\epsilon_s,B_s)$ is negligible.
The QDT expansion approaches 
the single-channel universal behavior \cite{gao09a} defined by
replacing $K^{c0}_l(\epsilon_s,B_s)$ in Eqs.~(\ref{eq:qdtexpB4}) and
(\ref{eq:qdtexpB5}) with its zero-energy value of $K^{c0}_l(B_s)$.
In such cases, multichannel scattering behaves the
same as single channel with a tunable (generalized) scattering length
not only at the threshold, but over a range of energies
around the threshold with an energy dependence determined
primary by the van der Waals interaction. 
This is illustrated in Figs.~\ref{fig:sbn1}
and \ref{fig:sbn2} using a broad resonance of $^6$Li-$^6$Li.
Narrow resonances behaves very differently with
a much more complex energy dependence that is determined
both by the properties of the resonance and by the
van der Waals interaction.
They change scattering in a narrow range of
energies around the resonance. Away from it, atomic
interaction evolves towards a single channel universal 
behavior determined not by $K^{c0}_l(B_s)$, but
by $K^{c0}_{\text{bg}l}$ or equivalently the
background (generalized) scattering length 
$\widetilde{a}_{\text{bg}l}$, with
\begin{equation}
K^{(D)}_l \sim -\bar{a}_{sl}k_s^{2l+1}\left[
	(-1)^l+\frac{(2l+3)(2l-1)-K^{c0}_{\text{bg}l}\epsilon_s}
	{(2l+3)(2l-1)K^{c0}_{\text{bg}l}+\epsilon_s+w_l\epsilon_s^2}\right] \;.
\label{eq:KDlbg}	
\end{equation}
Figures~\ref{fig:sbn1} and \ref{fig:sbn2} contain illustrations
of narrow-resonance behavior using an example from $^6$Li-$^6$Li
scattering.
Further conceptual understanding of the differences between broad
and narrow resonances can be found in the next section, in
connections with the generalized effective range expansion
and examples for infinite and zero (generalized) scattering lengths.

\begin{figure}
\scalebox{0.3}{\includegraphics{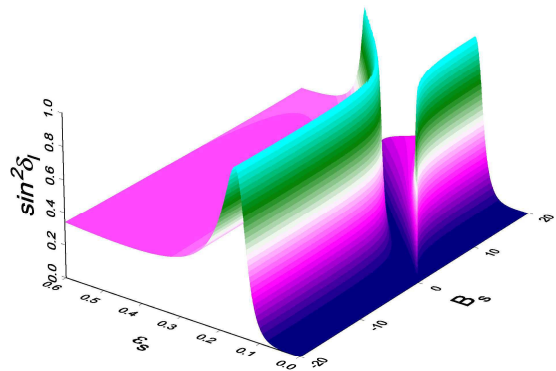}}
\caption{(Color online) A plot of $\sin^2\delta_l$ vs $\epsilon_s$ 
and $B_s$ for the $p$
wave, with parameters $K^{c0}_{\text{bg}l}=-0.03$ and $g_{\text{res}}=0.02$.
It illustrates the avoided crossing between a background shape resonance
located in the threshold region and a narrow Feshbach/shape resonance.
\label{fig:px}}
\end{figure}
As an illustration of the breadth of the physics 
contained in the QDT expansion for a magnetic Feshbach resonance,
Figure~\ref{fig:px} shows its description of an avoided
crossing between a narrow $p$ wave Feshbach/shape resonance and
a background $p$ wave shape resonance in the threshold region.
It is an example of the coupling of a bound state to
a highly ``structured'' continuum, and is used here to emphasize 
that the \textit{only} restriction on the applicability
of the QDT expansion is $\epsilon_s\ll \epsilon_{scl}$ \cite{gao09a}.

For nonzero partial waves, a Feshbach resonance above the threshold 
manifests itself as a Feshbach/shape resonance.
The qualitative characteristics of such a resonance, such as its
position and width, are
contained within the QDT expansion, in a manner similar to
the case of a single channel \cite{gao09a}. 
We defer their discussions to a following paper,
since they need to be combined with the binding
energy of a Feshbach molecule to give a complete picture of
the evolution of a resonance across the threshold.

\section{Sample applications}
\label{sec:apps}

\subsection{The generalized effective range expansion around a
magnetic Feshbach resonance}
\label{sec:tger}

One of the ways to understand some of the physics contained
in the QDT expansion for atomic interaction around a magnetic
Feshbach resonance is through the generalized effective range
expansion contained within it.
As in paper I \cite{gao09a}, the QDT expansion, given by 
Eqs.~(\ref{eq:qdtexpB1})-(\ref{eq:qdtexpB5}),
can be \textit{approximated} by an generalized effective range expansion
\begin{equation}
k^{2l+1}\cot(\delta_l-\delta^{(B)}_l)=-\frac{1}{\widetilde{a}_l}
	+\frac{1}{2}\widetilde{r}_{el}k^2 +O(k^4\ln k) \;,
\label{eq:gere}	
\end{equation}
where $\delta^{(B)}_l=-\pi(\nu-\nu_0)$ is approximated by
Eq.~(\ref{eq:qdtexpB2}).
It reduces to the standard effective range expansion \cite{sch47,bla49,bet49} 
for $l=0$, and serves to define the generalized scattering length
and the generalized effective range for other $l$ \cite{gao09a}.
For scattering around a magnetic Feshbach resonance,
both the (generalized) scattering length and the (generalized)
effective range become magnetic-field dependent. 
The (generalized) scattering length, $\widetilde{a}_l$, is tuned by
the magnetic field according to Eq.~(\ref{eq:gaBold}).
From the QDT expansion, it is straightforward to show
that the (generalized) effective range is given by
\begin{widetext}
\begin{eqnarray}
\widetilde{r}_{el}(B) &=& -\frac{2\bar{a}_{l}\beta_6^2}{(2l+3)(2l-1)[\widetilde{a}_{l}(B)]^2}
	\left[1+\left((-1)^l-\frac{\widetilde{a}_{l}(B)}{\bar{a}_{l}}\right)^2\right] \nonumber\\
	& &-\left(\frac{\hbar^2}{\mu \widetilde{a}_{\text{bg}l}\delta\mu_l\Delta_{Bl}}\right)
	\left(\frac{\Delta_{Bl}}{B-B_{0l}-\Delta_{Bl}}\right)^2 \;,
\label{eq:gr1} \\
	&=& -\frac{2\bar{a}_{l}\beta_6^2}{(2l+3)(2l-1)[\widetilde{a}_{l}(B)]^2}
	\left[1+\left((-1)^l-\frac{\widetilde{a}_{l}(B)}{\bar{a}_{l}}\right)^2\right] \nonumber\\
	& &+\left(\frac{1}{\zeta_{\text{res}}}\right)
	\frac{2\beta_6^2}{(2l+3)(2l-1)\bar{a}_{l}}
	\left(\frac{\Delta_{Bl}}{B-B_{0l}-\Delta_{Bl}}\right)^2 \;,
\label{eq:gr2} \\
	&=& -\frac{2\bar{a}_{l}\beta_6^2}{(2l+3)(2l-1)[\widetilde{a}_{l}(B)]^2}
	\left[1+\left((-1)^l-\frac{\widetilde{a}_{l}(B)}{\bar{a}_{l}}\right)^2\right] \nonumber\\
	& &+\left(\frac{1}{\zeta_{\text{res}}}\right)
	\frac{2\bar{a}_{l}\beta_6^2}{(2l+3)(2l-1)[\widetilde{a}_{l}(B)]^2}
	\left[\frac{\widetilde{a}_{l}(B)-\widetilde{a}_{\text{bg}l}}{\bar{a}_{l}}\right]^2 \;.
\label{eq:gr3}	
\end{eqnarray}
\end{widetext}
It consists of two terms. The first corresponds to the single-channel universal
behavior of paper I \cite{gao09a} and is due to the long-range van der Waals interaction.
The second term is due to the energy dependence of the effective
$K^{c0}_l$ that comes from the coupling to a bound state in closed channels.
It is given here in three different forms with distinctive insights.

In the broad-resonance limit of
$|\zeta_{\text{res}}|\rightarrow\infty$, the energy dependence of the effective
$K^{c0}_l$ is negligible, and the the result reduces to the single-channel 
universal behavior \cite{gao09a}, in which the (generalized) effective range is uniquely
determined by the (generalized) scattering length, $\widetilde{a}_{l}(B)$, independent
of the other details of the resonance. 
Within a narrow resonance, the (generalized) effective is changed
substantially from the universal behavior, which is one way to
understand its substantially different near-threshold behavior
as illustrated in Figs.~\ref{fig:sbn1} and \ref{fig:sbn2}.
This change, as characterized by the second terms in Eqs.~(\ref{eq:gr1})-(\ref{eq:gr3}), 
depends sensitively both on the location within the resonance and
on specific characteristics of the resonance, which are described,
for instance, by the $\zeta_{\text{res}}$ and $\widetilde{a}_{\text{bg}l}$ 
parameters in Eq.~(\ref{eq:gr3}).
Away from the resonance, namely for $|B-B_{0l}|\gg|\Delta_{Bl}|$, the second 
term goes away and the (generalized) effective range evolves back
towards a single-channel universal result \cite{gao09a} determined 
by the (generalized) background scattering length, $\widetilde{a}_{\text{bg}l}$.

It is useful to note that the contribution to the (generalized)
effective range due to the coupling to a bound state in closed 
channels, the second term, is always negative.
It comes from the constraint of $\Gamma^c>0$ discussed earlier,
which translates into $\delta\mu_l\widetilde{a}_{\text{bg}l}\Delta_{Bl}>0$
for parameters used in Eq.~(\ref{eq:gr1}), and into  
$\zeta_{\text{res}}>0$ for $l=0$ and $\zeta_{\text{res}}<0$ for 
all other partial waves, for parameters used in Eqs.~(\ref{eq:gr2})
and (\ref{eq:gr3}) (see Sec.~\ref{sec:scaledpara}). 
The implication is that $\widetilde{r}_{el}(B)$ is always negative
for $l>0$ as its corresponding first term is also always 
negative \cite{gao09a}.
For the $s$ wave, the two terms are always of opposite
signs and the end result can be either positive or negative.

Specializing to the case of $s$ wave, the second term in Eq.~(\ref{eq:gr1}) 
is the expression for the effective range that is adopted by
Zinner and Thogersen \cite{zin09} in
their investigation of Bose-Einstein condensate (BEC)
around a narrow Feshbach resonance. It comes from the work of 
Bruun \textit{et al.} \cite{bru05}, which ignores the effect of
van der Waals interaction.
Having only the second term in their theories 
means that that their results are applicable only for narrow resonances
($|\zeta_{\text{res}}|\ll 1$), and only in the resonance region where 
the scattering length differs substantially 
from the background scattering length.

We note that while the generalized effective range expansion
is useful both as a connection to previous theories and 
for studies of dilute quantum gases at ultracold temperatures 
\cite{fu03b,col07,zin09}, it has its limitations.
Similar to the case of a single channel \cite{gao09a}, it fails
around $\widetilde{a}_{l}(B)=0$, and has generally a much
more limited range of applicability compared to
the QDT expansion, from which it is derived.

\subsection{Examples of infinite and zero (generalized) scattering lengths}

Both for the purpose of illustrating explicit energy dependences
contained in the QDT expansion for magnetic Feshbach resonances,
and to facilitate future applications, we give here explicit
QDT expansion for two special cases of interest in cold-atom
physics. One is the case of infinite (generalized) scattering length,
the so-called unitary limit. The other is the case of zero
(generalized) scattering length. 

For infinite (generalized) scattering length, which occurs at $B=B_{0l}$, 
or equivalently, at $B_s=0$,  
the QDT expansion for $K^{(D)}_l$, Eqs.~(\ref{eq:qdtexpB3})
and (\ref{eq:qdtexpB5}), becomes
\begin{widetext}
\begin{equation}
K^{(D)}_l = \bar{a}_{sl}k_s^{2l-1}
	\frac{(2l+3)(2l-1)g_{\text{res}}-\epsilon_s
	\left[(2l+3)(2l-1)-K^{c0}_{\text{bg}l}\epsilon_s\right]}
	{(2l+3)(2l-1)K^{c0}_{\text{bg}l}+\epsilon_s+w_l\epsilon_s^2
	-g_{\text{res}}(1+w_l\epsilon_s)}
	-(-1)^l\bar{a}_{sl}k_s^{2l+1} \;.
\label{eq:KDlinf}	
\end{equation}
\end{widetext}
In the broad resonance limit of $|g_{\text{res}}|\rightarrow\infty$, 
it reduces to Eq.~(51) of paper I \cite{gao09a}.
For a narrow resonance with $|\zeta_{\text{res}}| \ll 1$,
scattering around the threshold follows that of an 
infinite (generalized) scattering length,
with $K^{(D)}_l\sim k_s^{2l-1}$ \cite{gao09a}, 
only in a very small energy 
range of $0<\epsilon_s\ll |g_{\text{res}}|$
around the threshold.
Outside this region, namely for energies $\epsilon_s\gg |g_{\text{res}}|$,
it evolves into the background scattering described by 
Eq.~(\ref{eq:KDlbg}), as discussed in the previous
section in a more general context.


Another special case of interest, where the effective range
expansion, including the generalized version of Sec.~\ref{sec:tger},
fails completely, is that of
zero (generalized) scattering length. It occurs at the magnetic
field $B=B_{0l}+\Delta_{Bl}$, or, equivalently at
$B_s = -[1+(-1)^lK^{c0}_{\mathrm{bg}l}]^{-1}$,
where Eqs.~(\ref{eq:qdtexpB3})
and (\ref{eq:qdtexpB5}) becomes
\begin{widetext}
\begin{equation}
K^{(D)}_l = \bar{a}_{sl}k_s^{2l+3}
	\frac{K^{c0}_{\text{bg}l}g_{\text{res}}(2+w_l\epsilon_s)
	-\left[(-1)^l+K^{c0}_{\text{bg}l}\right]\left\{(2l+3)(2l-1)K^{c0}_{\text{bg}l}+\epsilon_s+w_l\epsilon_s^2
	+(-1)^l\left[(2l+3)(2l-1)-K^{c0}_{\text{bg}l}\epsilon_s\right]\right\}}
	{K^{c0}_{\text{bg}l}g_{\text{res}}
	\left[(2l+3)(2l-1)K^{c0}_{\text{bg}l}-(-1)^l\epsilon_s-(-1)^lw_l\epsilon_s^2\right]
	+(-1)^l\left[(-1)^l+K^{c0}_{\text{bg}l}\right]\epsilon_s
	\left[(2l+3)(2l-1)K^{c0}_{\text{bg}l}+\epsilon_s+w_l\epsilon_s^2\right]} \;,
\label{eq:KDlzero}	
\end{equation}
\end{widetext}
In the broad resonance limit of $|g_{\text{res}}|\rightarrow\infty$, 
it reduces to Eq.~(50) of paper I.
For a narrow resonance with $|\zeta_{\text{res}}|\ll 1$,
it behaves as scattering of zero (generalized) scattering length, 
with $K^{(D)}_l\sim k_s^{2l+3}$ \cite{gao09a},
only in a small energy range
of $0<\epsilon_s\ll |g_{\text{res}}|$. 
Outside this range it becomes that determined
by the background scattering, as given by Eq.~(\ref{eq:KDlbg}).
Equations~(\ref{eq:KDlinf}) and (\ref{eq:KDlzero}) 
illustrate the kind of energy dependences
contained in the QDT expansion for magnetic Feshbach resonances,
and the complexity required to describe two types of
behaviors, one determined by $K^{c0}_l(B_s)$ or $\widetilde{a}_{l}(B)$
sufficiently close to the threshold, 
and one by $K^{c0}_{\text{bg}l}$ or $\widetilde{a}_{\text{bg}l}$
away from the resonance, 
and the evolution between the two.

\section{Conclusions}
\label{sec:conclude}

In conclusion, we have presented an analytic description of 
a magnetic Feshbach resonance in an arbitrary partial wave $l$ and the 
atomic scattering around it at ultracold temperatures.
It is derived by showing, in a very general context,
that a multichannel problem below the second threshold,
all the way through the bound spectrum, is
equivalent to an effective single-channel problem with
a generally energy- and partial-wave-dependent
short range parameter. The relative significance of
this energy dependence, in comparison with those induced
by the long-range interaction, leads to the classification
of Feshbach resonances of arbitrary $l$ into broad
and narrow resonances, with vastly different scattering
characteristics around the threshold.

We have shown that, except for the special case of 
$K^{c0}_{\text{bg}l}=0$ (corresponding to $\widetilde{a}_{\text{bg}l}=\infty$,
and discussed further in Appendix~\ref{sec:altpara}),
a magnetic Feshbach resonance of arbitrary $l$ can be parametrized
in a similar fashion as an $s$ wave Feshbach resonance \cite{moe95,koh06,chi10},
in terms of five parameters, which can be 
either $B_{0l}$, $\widetilde{a}_{\text{bg}l}$, $\Delta_{Bl}$, 
$\delta\mu_l$ and $s_E$ (or $C_6$), or
$B_{0l}$, $K^{c0}_{\text{bg}l}$, $g_{\text{res}}$, 
$d_{Bl}$, and $s_E$ (or $C_6$).
These parameters, together with the QDT expansion, give accurate
analytic descriptions of atomic interactions around a 
magnetic Feshbach resonance,
not only of the scattering properties presented here, but also
of the binding energies of a Feshbach molecule and of scattering
at negative energies \cite{gao08a}, 
to be presented in a following publication.
Such descriptions can now be incorporated into theories of atomic interaction
in an optical lattice \cite{hal10}, using, e.g., the multiscale
QDT of Ref.~\cite{che07}, and theories of few-atom and
many-atom systems around a Feshbach resonance, especially
around a resonance that is not broad 
(See, e.g., Refs.~\cite{fu03b,pet04b,sto05,szy05,col07,zin09}).

Accurate determinations of Feshbach parameters will generally 
require a combination of theoretical and experimental efforts, 
as have been done previously for the $s$ wave \cite{chi10}.
The derivation of the parametrization, as given in
Sec.~\ref{sec:reduce}-\ref{sec:para} and Appendix~\ref{sec:altpara}, 
also constitutes
an outline of a theory for these parameters and how they
can be computed from the MQDT formulation  
for atomic interaction in a magnetic field \cite{gao05a,han09}.
The detailed implementation of the theory and 
results for specific systems will be presented elsewhere.

It should be clear that while the focus of this article
is on magnetic Feshbach resonances, many of the concepts are much
more generally applicable. In particular, the theoretical development
followed here provides a very general 
methodology on how analytic descriptions of certain aspects of 
a multichannel problem may be developed and understood.
The theory is also a necessary step towards resolving one
remaining difficulty in the analytic description of ultracold
atomic interaction, which is to efficiently incorporate
the weak magnetic dipole-dipole and second-order spin-orbit
interactions \cite{sto88,moe95,mie96,kot00,leo00}. 
Before one can describe how such anistropic interactions
can, e.g., couple a $d$ wave resonance into the 
$s$ wave \cite{chi10}, it is first necessary to efficiently
characterize the resonances without such coupling.

\begin{acknowledgments}
This work was partially supported by the NSF through a grant for 
ITAMP at Harvard University and Smithsonian Astrophysical Observatory.
The work at Toledo was supported in part by the NSF under 
the Grant No. PHY-0758042.
\end{acknowledgments}

\appendix
\section{Reduction of an $N$ channel bound state problem to an effective
	$N_a<N$ channel bound state problem}
\label{sec:reduceDetail}

In MQDT for $-1/r^n$ type of potentials with $n>2$ \cite{gao05a}, 
the bound states 
energies for an $N$ channel problem is given generally by the solutions of
\begin{equation}
\det(\chi^c - K^{c}) = 0 \;,
\end{equation}
where $K^c$ is an $N\times N$ real and symmetric matrix, 
and $\chi^c$ is an $N\times N$ diagonal matrix with elements
$\chi^{c(n_i)}_l(\epsilon_{si})$.

Separating the $N$ channels into $N_a$ ``$a$'' channels and $N_c=N-N_a$
``$c$'' channels, the $K^c$ matrix can be written in a partitioned 
form as
\begin{equation}
K^c = \left(
\begin{array}{cc}
K^c_{aa} & K^c_{ac} \\
K^c_{ca} & K^c_{cc}
\end{array}
\right) \;,
\end{equation}
where $K^c_{aa}$ is a $N_a\times N_a$ submatrix of $K^c$,
$K^c_{cc}$ is a $N_c\times N_c$ submatrix,
$K^c_{ac}$ is a $N_a\times N_c$ submatrix, and
$K^c_{ca}$ is a $N_c\times N_a$ submatrix.
From $\det(xy) = \det(x)\det(y)$, we can write
\begin{eqnarray}
\lefteqn{\det(\chi^c - K^{c}) = 
	\det\left(
	\begin{array}{cc}
	\chi^c_{aa}-K^c_{aa} & K^c_{ac} \\
	K^c_{ca} &  \chi^c_{cc}-K^c_{cc}
	\end{array}
	\right) 
} \nonumber\\
&=& 
\det\left[A\left(
\begin{array}{cc}
\chi^c_{aa}-K^c_{aa} & K^c_{ac} \\
K^c_{ca} &  \chi^c_{cc}-K^c_{cc}
\end{array}
\right)A^{-1}B\right] \nonumber\\
&=& 
\det(A)\det\left[\left(
\begin{array}{cc}
\chi^c_{aa}-K^c_{aa} & K^c_{ac} \\
K^c_{ca} &  \chi^c_{cc}-K^c_{cc}
\end{array}
\right)A^{-1}B\right] 
\;.
\end{eqnarray}
Here $A$ is an arbitrary nonsingular matrix, and $B$ is an 
arbitrary matrix with $\det(B)=1$. Choosing
\begin{equation}
A = \left(
\begin{array}{cc}
I & 0 \\
0 &  \chi^c_{cc}-K^c_{cc}
\end{array}
\right) \;,
\end{equation}
and 
\begin{equation}
B = \left(
\begin{array}{cc}
I & 0 \\
-K^c_{ca} &  I
\end{array}
\right) \;,
\end{equation}
where $I$ represents an unit matrix,
we obtain
\begin{equation}
\det(\chi^c - K^{c}) = \det(\chi^c_{cc} - K^{c}_{cc})\det(\chi^c_{aa} - K^{c}_{\mathrm{eff}})\;,
\label{eq:detN}
\end{equation}
where
\begin{equation}
K^c_{\text{eff}} = K^{c}_{aa}+K^{c}_{ac}(\chi^c_{cc} - K^{c}_{cc})^{-1}K^{c}_{ca} \;.
\label{eq:Kceffg}
\end{equation}

Equation~(\ref{eq:detN}) means that if the channels ``$a$'' and channels ``$c$'' are not
coupled, namely $K^c_{ac}=0$, the bound states separate into
two sets, one for channels ``$a$'', given by $\det(\chi^c_{aa} - K^{c}_{aa})=0$,
and one for channels ``$c$'', given by $\det(\chi^c_{cc} - K^{c}_{cc})=0$.
(Since $K^c$, like any other $K$ matrix, is real and symmetric, $K^c_{ac}=0$
also implies $K^c_{ca}=(K^c_{ac})^{T}=0$.)
If they are coupled, which is the case that we are interested in,
the solutions of $\det(\chi^c_{cc} - K^{c}_{cc})=0$ are no longer 
solutions of $\det(\chi^c - K^{c})=0$.
This can be proven by taking the limit of $E\rightarrow \bar{E}$
in Eq.~(\ref{eq:detN}), where $\bar{E}$ is one of the solutions of
$\det(\chi^c_{cc} - K^{c}_{cc})=0$, namely one of the bare resonance
energies. 
Thus in the coupled case, all bound state energies are given by the 
solutions of
\begin{equation}
\det(\chi^c_{aa} - K^{c}_{\text{eff}}) = 0 \;,
\label{eq:bspg}
\end{equation}
with the effective $K^c$ matrix being given by Eq.~(\ref{eq:Kceffg}).

This procedure can in principle reduce an $N$ channel bound state problem to an effective
$N_a<N$ channel problem with $N_a$ being an arbitrary number smaller than $N$.
In reality, the choice of $N_a$ is of course determined by the underlying physics.
While we are using $N_a=1$ in this paper, in which case Eq.~(\ref{eq:bspg})
reduces to Eq.~(\ref{eq:bsp}), other choices are possible, and are
likely to be important in future treatments that incorporate
the magnetic dipole-dipole and second-order spin-orbit 
interactions \cite{sto88,moe95,mie96,kot00,leo00}.

\section{An alternative parametrization of magnetic Feshbach resonances and 
	the special case of infinite (generalized) background scattering length}
\label{sec:altpara}	

As stated in the main text, the parametrization adopted,
Eqs.~(\ref{eq:Kc0Fesh}) for the $K^{c0}_l(\epsilon,B)$ and the corresponding 
Eq.~(\ref{eq:gaBold}) for the generalized scattering length,
have the limitation that they fail for $K^{c0}_{\text{bg}l}=0$,
corresponding to an infinite (generalized) background scattering length
$\widetilde{a}_{\text{bg}l}=\infty$.
We show in this appendix that this is not an intrinsic
difficulty of the theory, but due simply to the desire of
using parameters that have more direct experimental interpretations.
There are parametrizations
of $K^{c0}_l$ and $\widetilde{a}_{l}$ that would work for arbitrary
background scattering lengths,
provided that we are willing to sacrifice using $B_{0l}$.

It is not surprising that \textit{any} parametrization based on
$B_{0l}$ would have difficult at $K^{c0}_{\text{bg}l}= 0$ 
($\widetilde{a}_{\text{bg}l}=\infty$), corresponding
to having a bound or quasibound background state right at the threshold.
The interaction of this background state and the ``bare'' Feshbach
state is such that $B_{0l}$ no longer exists, meaning that there can
never be, in this case, a (coupled) bound state right at the 
threshold due to avoided crossing.
This is reflected in the fact that the effective $K^{c0}_l$, 
given by Eq.~(\ref{eq:Kc0pag}), does not have
a solution for $K^{c0}_{l}(\epsilon=0,B_{0l})=0$ in the
special case of $K^{c0}_{\text{bg}l}= 0$.

This difficulty can be overcome by
expanding $\bar{\epsilon}_l(B)$ in Eq.~(\ref{eq:Kc0pag}) around
$\bar{B}_{0l}$, determined by 
$\bar{\epsilon}_l(\bar{B}_{0l})=0$, which is the magnetic field at 
which the ``bare'' Feshbach resonance is crossing the threshold.
This gives us $\bar{\epsilon}_l(B)\approx \delta\bar{\mu}_l(B-\bar{B}_{0l})$,
where $\delta\bar{\mu}_l = \left.d\bar{\epsilon}_l(B)/dB\right|_{B=\bar{B}_{0l}}$.
Equation~(\ref{eq:Kc0pag}) now becomes
\begin{equation}
K^{c0}_{l}(\epsilon,B) = 
	K^{c0}_{\mathrm{bg}l} -\frac{\Gamma^{c0}_l/2}{\epsilon-\delta\bar{\mu}_l(B-\bar{B}_{0l})-f_{El}}
	\;.
\label{eq:Kc0la}	
\end{equation}
It is a parametrization with four parameters,
$K^{c0}_{\text{bg}l}$, $\bar{B}_{0l}$, $\delta\bar{\mu}_l$, and $\Gamma^{c0}_l$.
[The $f_{El}$ is not an independent parameter and is still given by
Eq.~(\ref{eq:fEl}).]
The corresponding parametrization of $\widetilde{a}_{l}$ is
\begin{equation}
\widetilde{a}_{l}(B) = \bar{a}_l
	\frac{[1+(-1)^lK^{c0}_{\mathrm{bg}l}]
	(B-\bar{B}_{0l}+\bar{f}_{Bl})+(-1)^l\Gamma^{c0}_{Bl}/2}
	{K^{c0}_{\mathrm{bg}l}
	(B-\bar{B}_{0l}+\bar{f}_{Bl})+\Gamma^{c0}_{Bl}/2} \;.
\label{eq:gaBa}	
\end{equation}
where $\bar{f}_{Bl}=f_{El}/\delta\bar{\mu}_l$
and $\Gamma^{c0}_{Bl}=\Gamma^{c0}_l/\delta\bar{\mu}_l$.
Equations~(\ref{eq:Kc0la}) and (\ref{eq:gaBa}) 
work for both infinite and zero (generalized) background scattering lengths.
For example, in the case of $K^{c0}_{\text{bg}l}= 0$ 
($\widetilde{a}_{\text{bg}l}= \infty$), Eq.~(\ref{eq:gaBa}) becomes
\begin{equation}
\widetilde{a}_{l}(B) = \bar{a}_l
	\left[(-1)^l+\tan(\pi\nu_0/2)
	+\frac{B-\bar{B}_{0l}}{\Gamma^{c0}_{Bl}/2}\right] \;.
\label{eq:gaBbginf}	
\end{equation}
The disadvantage of this parametrization is that the parameter
$\bar{B}_{0l}$ does not have as direct of an experimental interpretation
as $B_{0l}$. It is more of a theoretical concept corresponding to
the magnetic field at which a ``bare'' Feshbach resonance is crossing
the threshold. The utility of this parametrization is, however,
beyond conceptual completeness. Depending on the range of magnetic
field of interest, it can be the preferred parametrization
for special cases with large background scattering lengths 
(see, e.g., Ref.~\cite{mar04}), and can be
used with the QDT expansion in a similar manner as the one adopted
in the main text. 

\bibliography{bgao,twobody,Fesh,manybody,fewbody,reducedDim}

\begin{thebibliography}{58}%
\makeatletter
\providecommand \@ifxundefined [1]{%
 \@ifx{#1\undefined}
}%
\providecommand \@ifnum [1]{%
 \ifnum #1\expandafter \@firstoftwo
 \else \expandafter \@secondoftwo
 \fi
}%
\providecommand \@ifx [1]{%
 \ifx #1\expandafter \@firstoftwo
 \else \expandafter \@secondoftwo
 \fi
}%
\providecommand \natexlab [1]{#1}%
\providecommand \enquote  [1]{``#1''}%
\providecommand \bibnamefont  [1]{#1}%
\providecommand \bibfnamefont [1]{#1}%
\providecommand \citenamefont [1]{#1}%
\providecommand \href@noop [0]{\@secondoftwo}%
\providecommand \href [0]{\begingroup \@sanitize@url \@href}%
\providecommand \@href[1]{\@@startlink{#1}\@@href}%
\providecommand \@@href[1]{\endgroup#1\@@endlink}%
\providecommand \@sanitize@url [0]{\catcode `\\12\catcode `\$12\catcode
  `\&12\catcode `\#12\catcode `\^12\catcode `\_12\catcode `\%12\relax}%
\providecommand \@@startlink[1]{}%
\providecommand \@@endlink[0]{}%
\providecommand \url  [0]{\begingroup\@sanitize@url \@url }%
\providecommand \@url [1]{\endgroup\@href {#1}{\urlprefix }}%
\providecommand \urlprefix  [0]{URL }%
\providecommand \Eprint [0]{\href }%
\providecommand \doibase [0]{http://dx.doi.org/}%
\providecommand \selectlanguage [0]{\@gobble}%
\providecommand \bibinfo  [0]{\@secondoftwo}%
\providecommand \bibfield  [0]{\@secondoftwo}%
\providecommand \translation [1]{[#1]}%
\providecommand \BibitemOpen [0]{}%
\providecommand \bibitemStop [0]{}%
\providecommand \bibitemNoStop [0]{.\EOS\space}%
\providecommand \EOS [0]{\spacefactor3000\relax}%
\providecommand \BibitemShut  [1]{\csname bibitem#1\endcsname}%
\let\auto@bib@innerbib\@empty
\bibitem [{\citenamefont {Dalfovo}\ \emph {et~al.}(1999)\citenamefont
  {Dalfovo}, \citenamefont {Giorgini}, \citenamefont {Pitaevskii},\ and\
  \citenamefont {Stringari}}]{dal99}%
  \BibitemOpen
  \bibfield  {author} {\bibinfo {author} {\bibfnamefont {F.}~\bibnamefont
  {Dalfovo}}, \bibinfo {author} {\bibfnamefont {S.}~\bibnamefont {Giorgini}},
  \bibinfo {author} {\bibfnamefont {L.~P.}\ \bibnamefont {Pitaevskii}}, \ and\
  \bibinfo {author} {\bibfnamefont {S.}~\bibnamefont {Stringari}},\ }\href@noop
  {} {\bibfield  {journal} {\bibinfo  {journal} {Rev. Mod. Phys.}\ }\textbf
  {\bibinfo {volume} {71}},\ \bibinfo {pages} {463} (\bibinfo {year}
  {1999})}\BibitemShut {NoStop}%
\bibitem [{\citenamefont {Schwinger}(1947)}]{sch47}%
  \BibitemOpen
  \bibfield  {author} {\bibinfo {author} {\bibfnamefont {J.}~\bibnamefont
  {Schwinger}},\ }\href {\doibase 10.1103/PhysRev.72.738} {\bibfield  {journal}
  {\bibinfo  {journal} {Phys. Rev.}\ }\textbf {\bibinfo {volume} {72}},\
  \bibinfo {pages} {738} (\bibinfo {year} {1947})}\BibitemShut {NoStop}%
\bibitem [{\citenamefont {Blatt}\ and\ \citenamefont {Jackson}(1949)}]{bla49}%
  \BibitemOpen
  \bibfield  {author} {\bibinfo {author} {\bibfnamefont {J.~M.}\ \bibnamefont
  {Blatt}}\ and\ \bibinfo {author} {\bibfnamefont {D.~J.}\ \bibnamefont
  {Jackson}},\ }\href@noop {} {\bibfield  {journal} {\bibinfo  {journal} {Phys.
  Rev.}\ }\textbf {\bibinfo {volume} {76}},\ \bibinfo {pages} {18} (\bibinfo
  {year} {1949})}\BibitemShut {NoStop}%
\bibitem [{\citenamefont {Bethe}(1949)}]{bet49}%
  \BibitemOpen
  \bibfield  {author} {\bibinfo {author} {\bibfnamefont {H.~A.}\ \bibnamefont
  {Bethe}},\ }\href {\doibase 10.1103/PhysRev.76.38} {\bibfield  {journal}
  {\bibinfo  {journal} {Phys. Rev.}\ }\textbf {\bibinfo {volume} {76}},\
  \bibinfo {pages} {38} (\bibinfo {year} {1949})}\BibitemShut {NoStop}%
\bibitem [{\citenamefont {Braaten}\ and\ \citenamefont {Hammer}(2006)}]{bra06}%
  \BibitemOpen
  \bibfield  {author} {\bibinfo {author} {\bibfnamefont {E.}~\bibnamefont
  {Braaten}}\ and\ \bibinfo {author} {\bibfnamefont {H.-W.}\ \bibnamefont
  {Hammer}},\ }\href {\doibase DOI: 10.1016/j.physrep.2006.03.001} {\bibfield
  {journal} {\bibinfo  {journal} {Physics Reports}\ }\textbf {\bibinfo {volume}
  {428}},\ \bibinfo {pages} {259 } (\bibinfo {year} {2006})}\BibitemShut
  {NoStop}%
\bibitem [{\citenamefont {Greene}(2010)}]{gre10}%
  \BibitemOpen
  \bibfield  {author} {\bibinfo {author} {\bibfnamefont {C.~H.}\ \bibnamefont
  {Greene}},\ }\href@noop {} {\bibfield  {journal} {\bibinfo  {journal}
  {Physics Today}\ }\textbf {\bibinfo {volume} {63}},\ \bibinfo {pages} {40}
  (\bibinfo {year} {2010})}\BibitemShut {NoStop}%
\bibitem [{\citenamefont {Gao}(2009)}]{gao09a}%
  \BibitemOpen
  \bibfield  {author} {\bibinfo {author} {\bibfnamefont {B.}~\bibnamefont
  {Gao}},\ }\href@noop {} {\bibfield  {journal} {\bibinfo  {journal} {Phys.
  Rev. A}\ }\textbf {\bibinfo {volume} {80}},\ \bibinfo {eid} {012702}
  (\bibinfo {year} {2009})}\BibitemShut {NoStop}%
\bibitem [{\citenamefont {Gao}(1998)}]{gao98b}%
  \BibitemOpen
  \bibfield  {author} {\bibinfo {author} {\bibfnamefont {B.}~\bibnamefont
  {Gao}},\ }\href@noop {} {\bibfield  {journal} {\bibinfo  {journal} {Phys.
  Rev. A}\ }\textbf {\bibinfo {volume} {58}},\ \bibinfo {pages} {4222}
  (\bibinfo {year} {1998})}\BibitemShut {NoStop}%
\bibitem [{\citenamefont {Gao}(2001)}]{gao01}%
  \BibitemOpen
  \bibfield  {author} {\bibinfo {author} {\bibfnamefont {B.}~\bibnamefont
  {Gao}},\ }\href@noop {} {\bibfield  {journal} {\bibinfo  {journal} {Phys.
  Rev. A}\ }\textbf {\bibinfo {volume} {64}},\ \bibinfo {pages} {010701(R)}
  (\bibinfo {year} {2001})}\BibitemShut {NoStop}%
\bibitem [{\citenamefont {Gao}(2008)}]{gao08a}%
  \BibitemOpen
  \bibfield  {author} {\bibinfo {author} {\bibfnamefont {B.}~\bibnamefont
  {Gao}},\ }\href@noop {} {\bibfield  {journal} {\bibinfo  {journal} {Phys.
  Rev. A}\ }\textbf {\bibinfo {volume} {78}},\ \bibinfo {pages} {012702}
  (\bibinfo {year} {2008})}\BibitemShut {NoStop}%
\bibitem [{\citenamefont {Tiesinga}\ \emph {et~al.}(1993)\citenamefont
  {Tiesinga}, \citenamefont {Verhaar},\ and\ \citenamefont {Stoof}}]{tie93}%
  \BibitemOpen
  \bibfield  {author} {\bibinfo {author} {\bibfnamefont {E.}~\bibnamefont
  {Tiesinga}}, \bibinfo {author} {\bibfnamefont {B.~J.}\ \bibnamefont
  {Verhaar}}, \ and\ \bibinfo {author} {\bibfnamefont {H.~T.~C.}\ \bibnamefont
  {Stoof}},\ }\href@noop {} {\bibfield  {journal} {\bibinfo  {journal} {Phys.
  Rev. A}\ }\textbf {\bibinfo {volume} {47}},\ \bibinfo {pages} {4114}
  (\bibinfo {year} {1993})}\BibitemShut {NoStop}%
\bibitem [{\citenamefont {{K\"{o}hler}}\ \emph {et~al.}(2006)\citenamefont
  {{K\"{o}hler}}, \citenamefont {{G\'{o}ral}},\ and\ \citenamefont
  {Julienne}}]{koh06}%
  \BibitemOpen
  \bibfield  {author} {\bibinfo {author} {\bibfnamefont {T.}~\bibnamefont
  {{K\"{o}hler}}}, \bibinfo {author} {\bibfnamefont {K.}~\bibnamefont
  {{G\'{o}ral}}}, \ and\ \bibinfo {author} {\bibfnamefont {P.~S.}\ \bibnamefont
  {Julienne}},\ }\href@noop {} {\bibfield  {journal} {\bibinfo  {journal} {Rev.
  Mod. Phys.}\ }\textbf {\bibinfo {volume} {78}},\ \bibinfo {pages} {1311}
  (\bibinfo {year} {2006})}\BibitemShut {NoStop}%
\bibitem [{\citenamefont {Chin}\ \emph {et~al.}(2010)\citenamefont {Chin},
  \citenamefont {Grimm}, \citenamefont {Julienne},\ and\ \citenamefont
  {Tiesinga}}]{chi10}%
  \BibitemOpen
  \bibfield  {author} {\bibinfo {author} {\bibfnamefont {C.}~\bibnamefont
  {Chin}}, \bibinfo {author} {\bibfnamefont {R.}~\bibnamefont {Grimm}},
  \bibinfo {author} {\bibfnamefont {P.}~\bibnamefont {Julienne}}, \ and\
  \bibinfo {author} {\bibfnamefont {E.}~\bibnamefont {Tiesinga}},\ }\href
  {\doibase 10.1103/RevModPhys.82.1225} {\bibfield  {journal} {\bibinfo
  {journal} {Rev. Mod. Phys.}\ }\textbf {\bibinfo {volume} {82}},\ \bibinfo
  {pages} {1225} (\bibinfo {year} {2010})}\BibitemShut {NoStop}%
\bibitem [{\citenamefont {Stoll}\ and\ \citenamefont {K\"ohler}(2005)}]{sto05}%
  \BibitemOpen
  \bibfield  {author} {\bibinfo {author} {\bibfnamefont {M.}~\bibnamefont
  {Stoll}}\ and\ \bibinfo {author} {\bibfnamefont {T.}~\bibnamefont
  {K\"ohler}},\ }\href {\doibase 10.1103/PhysRevA.72.022714} {\bibfield
  {journal} {\bibinfo  {journal} {Phys. Rev. A}\ }\textbf {\bibinfo {volume}
  {72}},\ \bibinfo {pages} {022714} (\bibinfo {year} {2005})}\BibitemShut
  {NoStop}%
\bibitem [{\citenamefont {Simonucci}\ \emph {et~al.}(2005)\citenamefont
  {Simonucci}, \citenamefont {Pieri},\ and\ \citenamefont {Strinati}}]{sim05}%
  \BibitemOpen
  \bibfield  {author} {\bibinfo {author} {\bibfnamefont {S.}~\bibnamefont
  {Simonucci}}, \bibinfo {author} {\bibfnamefont {P.}~\bibnamefont {Pieri}}, \
  and\ \bibinfo {author} {\bibfnamefont {G.~C.}\ \bibnamefont {Strinati}},\
  }\href@noop {} {\bibfield  {journal} {\bibinfo  {journal} {Europhys. Lett.}\
  }\textbf {\bibinfo {volume} {69}},\ \bibinfo {pages} {713} (\bibinfo {year}
  {2005})}\BibitemShut {NoStop}%
\bibitem [{\citenamefont {Haller}\ \emph {et~al.}(2010)\citenamefont {Haller},
  \citenamefont {Mark}, \citenamefont {Hart}, \citenamefont {Danzl},
  \citenamefont {Reichs\"ollner}, \citenamefont {Melezhik}, \citenamefont
  {Schmelcher},\ and\ \citenamefont {N\"agerl}}]{hal10}%
  \BibitemOpen
  \bibfield  {author} {\bibinfo {author} {\bibfnamefont {E.}~\bibnamefont
  {Haller}}, \bibinfo {author} {\bibfnamefont {M.~J.}\ \bibnamefont {Mark}},
  \bibinfo {author} {\bibfnamefont {R.}~\bibnamefont {Hart}}, \bibinfo {author}
  {\bibfnamefont {J.~G.}\ \bibnamefont {Danzl}}, \bibinfo {author}
  {\bibfnamefont {L.}~\bibnamefont {Reichs\"ollner}}, \bibinfo {author}
  {\bibfnamefont {V.}~\bibnamefont {Melezhik}}, \bibinfo {author}
  {\bibfnamefont {P.}~\bibnamefont {Schmelcher}}, \ and\ \bibinfo {author}
  {\bibfnamefont {H.-C.}\ \bibnamefont {N\"agerl}},\ }\href {\doibase
  10.1103/PhysRevLett.104.153203} {\bibfield  {journal} {\bibinfo  {journal}
  {Phys. Rev. Lett.}\ }\textbf {\bibinfo {volume} {104}},\ \bibinfo {pages}
  {153203} (\bibinfo {year} {2010})}\BibitemShut {NoStop}%
\bibitem [{\citenamefont {Chen}\ and\ \citenamefont {Gao}(2007)}]{che07}%
  \BibitemOpen
  \bibfield  {author} {\bibinfo {author} {\bibfnamefont {Y.}~\bibnamefont
  {Chen}}\ and\ \bibinfo {author} {\bibfnamefont {B.}~\bibnamefont {Gao}},\
  }\href@noop {} {\bibfield  {journal} {\bibinfo  {journal} {Phys. Rev. A}\
  }\textbf {\bibinfo {volume} {75}},\ \bibinfo {pages} {053601} (\bibinfo
  {year} {2007})}\BibitemShut {NoStop}%
\bibitem [{\citenamefont {Zhang}\ \emph {et~al.}(2010)\citenamefont {Zhang},
  \citenamefont {Naidon},\ and\ \citenamefont {Ueda}}]{zha10}%
  \BibitemOpen
  \bibfield  {author} {\bibinfo {author} {\bibfnamefont {P.}~\bibnamefont
  {Zhang}}, \bibinfo {author} {\bibfnamefont {P.}~\bibnamefont {Naidon}}, \
  and\ \bibinfo {author} {\bibfnamefont {M.}~\bibnamefont {Ueda}},\ }\href
  {\doibase 10.1103/PhysRevA.82.062712} {\bibfield  {journal} {\bibinfo
  {journal} {Phys. Rev. A}\ }\textbf {\bibinfo {volume} {82}},\ \bibinfo
  {pages} {062712} (\bibinfo {year} {2010})}\BibitemShut {NoStop}%
\bibitem [{\citenamefont {Petrov}(2004)}]{pet04b}%
  \BibitemOpen
  \bibfield  {author} {\bibinfo {author} {\bibfnamefont {D.~S.}\ \bibnamefont
  {Petrov}},\ }\href {\doibase 10.1103/PhysRevLett.93.143201} {\bibfield
  {journal} {\bibinfo  {journal} {Phys. Rev. Lett.}\ }\textbf {\bibinfo
  {volume} {93}},\ \bibinfo {pages} {143201} (\bibinfo {year}
  {2004})}\BibitemShut {NoStop}%
\bibitem [{\citenamefont {Fu}\ \emph {et~al.}(2003)\citenamefont {Fu},
  \citenamefont {Wang},\ and\ \citenamefont {Gao}}]{fu03b}%
  \BibitemOpen
  \bibfield  {author} {\bibinfo {author} {\bibfnamefont {H.}~\bibnamefont
  {Fu}}, \bibinfo {author} {\bibfnamefont {Y.}~\bibnamefont {Wang}}, \ and\
  \bibinfo {author} {\bibfnamefont {B.}~\bibnamefont {Gao}},\ }\href@noop {}
  {\bibfield  {journal} {\bibinfo  {journal} {Phys. Rev. A}\ }\textbf {\bibinfo
  {volume} {67}},\ \bibinfo {pages} {053612} (\bibinfo {year}
  {2003})}\BibitemShut {NoStop}%
\bibitem [{\citenamefont {Szyma\ifmmode~\acute{n}\else \'{n}\fi{}ska}\ \emph
  {et~al.}(2005)\citenamefont {Szyma\ifmmode~\acute{n}\else \'{n}\fi{}ska},
  \citenamefont {G\'oral}, \citenamefont {K\"ohler},\ and\ \citenamefont
  {Burnett}}]{szy05}%
  \BibitemOpen
  \bibfield  {author} {\bibinfo {author} {\bibfnamefont {M.~H.}\ \bibnamefont
  {Szyma\ifmmode~\acute{n}\else \'{n}\fi{}ska}}, \bibinfo {author}
  {\bibfnamefont {K.}~\bibnamefont {G\'oral}}, \bibinfo {author} {\bibfnamefont
  {T.}~\bibnamefont {K\"ohler}}, \ and\ \bibinfo {author} {\bibfnamefont
  {K.}~\bibnamefont {Burnett}},\ }\href {\doibase 10.1103/PhysRevA.72.013610}
  {\bibfield  {journal} {\bibinfo  {journal} {Phys. Rev. A}\ }\textbf {\bibinfo
  {volume} {72}},\ \bibinfo {pages} {013610} (\bibinfo {year}
  {2005})}\BibitemShut {NoStop}%
\bibitem [{\citenamefont {Collin}\ \emph {et~al.}(2007)\citenamefont {Collin},
  \citenamefont {Massignan},\ and\ \citenamefont {Pethick}}]{col07}%
  \BibitemOpen
  \bibfield  {author} {\bibinfo {author} {\bibfnamefont {A.}~\bibnamefont
  {Collin}}, \bibinfo {author} {\bibfnamefont {P.}~\bibnamefont {Massignan}}, \
  and\ \bibinfo {author} {\bibfnamefont {C.~J.}\ \bibnamefont {Pethick}},\
  }\href {\doibase 10.1103/PhysRevA.75.013615} {\bibfield  {journal} {\bibinfo
  {journal} {Phys. Rev. A}\ }\textbf {\bibinfo {volume} {75}},\ \bibinfo {eid}
  {013615} (\bibinfo {year} {2007})}\BibitemShut {NoStop}%
\bibitem [{\citenamefont {Zinner}\ and\ \citenamefont
  {Thogersen}(2009)}]{zin09}%
  \BibitemOpen
  \bibfield  {author} {\bibinfo {author} {\bibfnamefont {N.~T.}\ \bibnamefont
  {Zinner}}\ and\ \bibinfo {author} {\bibfnamefont {M.}~\bibnamefont
  {Thogersen}},\ }\href {\doibase 10.1103/PhysRevA.80.023607} {\bibfield
  {journal} {\bibinfo  {journal} {Phys. Rev. A}\ }\textbf {\bibinfo {volume}
  {80}},\ \bibinfo {eid} {023607} (\bibinfo {year} {2009})}\BibitemShut
  {NoStop}%
\bibitem [{\citenamefont {Gao}\ \emph {et~al.}(2005)\citenamefont {Gao},
  \citenamefont {Tiesinga}, \citenamefont {Williams},\ and\ \citenamefont
  {Julienne}}]{gao05a}%
  \BibitemOpen
  \bibfield  {author} {\bibinfo {author} {\bibfnamefont {B.}~\bibnamefont
  {Gao}}, \bibinfo {author} {\bibfnamefont {E.}~\bibnamefont {Tiesinga}},
  \bibinfo {author} {\bibfnamefont {C.~J.}\ \bibnamefont {Williams}}, \ and\
  \bibinfo {author} {\bibfnamefont {P.~S.}\ \bibnamefont {Julienne}},\
  }\href@noop {} {\bibfield  {journal} {\bibinfo  {journal} {Phys. Rev. A}\
  }\textbf {\bibinfo {volume} {72}},\ \bibinfo {pages} {042719} (\bibinfo
  {year} {2005})}\BibitemShut {NoStop}%
\bibitem [{\citenamefont {Moerdijk}\ \emph {et~al.}(1995)\citenamefont
  {Moerdijk}, \citenamefont {Verhaar},\ and\ \citenamefont {Axelsson}}]{moe95}%
  \BibitemOpen
  \bibfield  {author} {\bibinfo {author} {\bibfnamefont {A.~J.}\ \bibnamefont
  {Moerdijk}}, \bibinfo {author} {\bibfnamefont {B.~J.}\ \bibnamefont
  {Verhaar}}, \ and\ \bibinfo {author} {\bibfnamefont {A.}~\bibnamefont
  {Axelsson}},\ }\href {\doibase 10.1103/PhysRevA.51.4852} {\bibfield
  {journal} {\bibinfo  {journal} {Phys. Rev. A}\ }\textbf {\bibinfo {volume}
  {51}},\ \bibinfo {pages} {4852} (\bibinfo {year} {1995})}\BibitemShut
  {NoStop}%
\bibitem [{\citenamefont {Flambaum}\ \emph {et~al.}(1999)\citenamefont
  {Flambaum}, \citenamefont {Gribakin},\ and\ \citenamefont
  {Harabati}}]{fla99}%
  \BibitemOpen
  \bibfield  {author} {\bibinfo {author} {\bibfnamefont {V.~V.}\ \bibnamefont
  {Flambaum}}, \bibinfo {author} {\bibfnamefont {G.~F.}\ \bibnamefont
  {Gribakin}}, \ and\ \bibinfo {author} {\bibfnamefont {C.}~\bibnamefont
  {Harabati}},\ }\href@noop {} {\bibfield  {journal} {\bibinfo  {journal}
  {Phys. Rev. A}\ }\textbf {\bibinfo {volume} {59}},\ \bibinfo {pages} {1998}
  (\bibinfo {year} {1999})}\BibitemShut {NoStop}%
\bibitem [{\citenamefont {Stoof}\ \emph {et~al.}(1988)\citenamefont {Stoof},
  \citenamefont {Koelman},\ and\ \citenamefont {Verhaar}}]{sto88}%
  \BibitemOpen
  \bibfield  {author} {\bibinfo {author} {\bibfnamefont {H.~T.~C.}\
  \bibnamefont {Stoof}}, \bibinfo {author} {\bibfnamefont {J.~M. V.~A.}\
  \bibnamefont {Koelman}}, \ and\ \bibinfo {author} {\bibfnamefont {B.~J.}\
  \bibnamefont {Verhaar}},\ }\href@noop {} {\bibfield  {journal} {\bibinfo
  {journal} {Phys. Rev. B}\ }\textbf {\bibinfo {volume} {38}},\ \bibinfo
  {pages} {4688} (\bibinfo {year} {1988})}\BibitemShut {NoStop}%
\bibitem [{\citenamefont {Mies}\ \emph {et~al.}(1996)\citenamefont {Mies},
  \citenamefont {Williams}, \citenamefont {Julienne},\ and\ \citenamefont
  {Krauss}}]{mie96}%
  \BibitemOpen
  \bibfield  {author} {\bibinfo {author} {\bibfnamefont {F.~H.}\ \bibnamefont
  {Mies}}, \bibinfo {author} {\bibfnamefont {C.~J.}\ \bibnamefont {Williams}},
  \bibinfo {author} {\bibfnamefont {P.~S.}\ \bibnamefont {Julienne}}, \ and\
  \bibinfo {author} {\bibfnamefont {M.}~\bibnamefont {Krauss}},\ }\href@noop {}
  {\bibfield  {journal} {\bibinfo  {journal} {J. Res. Natl. Inst. Stand.
  Technol.}\ }\textbf {\bibinfo {volume} {101}},\ \bibinfo {pages} {521}
  (\bibinfo {year} {1996})}\BibitemShut {NoStop}%
\bibitem [{\citenamefont {Kotochigova}\ \emph {et~al.}(2000)\citenamefont
  {Kotochigova}, \citenamefont {Tiesinga},\ and\ \citenamefont
  {Julienne}}]{kot00}%
  \BibitemOpen
  \bibfield  {author} {\bibinfo {author} {\bibfnamefont {S.}~\bibnamefont
  {Kotochigova}}, \bibinfo {author} {\bibfnamefont {E.}~\bibnamefont
  {Tiesinga}}, \ and\ \bibinfo {author} {\bibfnamefont {P.~S.}\ \bibnamefont
  {Julienne}},\ }\href {\doibase 10.1103/PhysRevA.63.012517} {\bibfield
  {journal} {\bibinfo  {journal} {Phys. Rev. A}\ }\textbf {\bibinfo {volume}
  {63}},\ \bibinfo {pages} {012517} (\bibinfo {year} {2000})}\BibitemShut
  {NoStop}%
\bibitem [{\citenamefont {Leo}\ \emph {et~al.}(2000)\citenamefont {Leo},
  \citenamefont {Williams},\ and\ \citenamefont {Julienne}}]{leo00}%
  \BibitemOpen
  \bibfield  {author} {\bibinfo {author} {\bibfnamefont {P.~J.}\ \bibnamefont
  {Leo}}, \bibinfo {author} {\bibfnamefont {C.~J.}\ \bibnamefont {Williams}}, \
  and\ \bibinfo {author} {\bibfnamefont {P.~S.}\ \bibnamefont {Julienne}},\
  }\href@noop {} {\bibfield  {journal} {\bibinfo  {journal} {Phys. Rev. Lett.}\
  }\textbf {\bibinfo {volume} {85}},\ \bibinfo {pages} {2721} (\bibinfo {year}
  {2000})}\BibitemShut {NoStop}%
\bibitem [{\citenamefont {Hanna}\ \emph {et~al.}(2009)\citenamefont {Hanna},
  \citenamefont {Tiesinga},\ and\ \citenamefont {Julienne}}]{han09}%
  \BibitemOpen
  \bibfield  {author} {\bibinfo {author} {\bibfnamefont {T.~M.}\ \bibnamefont
  {Hanna}}, \bibinfo {author} {\bibfnamefont {E.}~\bibnamefont {Tiesinga}}, \
  and\ \bibinfo {author} {\bibfnamefont {P.~S.}\ \bibnamefont {Julienne}},\
  }\href@noop {} {\bibfield  {journal} {\bibinfo  {journal} {Phys. Rev. A}\
  }\textbf {\bibinfo {volume} {79}},\ \bibinfo {eid} {040701} (\bibinfo {year}
  {2009})}\BibitemShut {NoStop}%
\bibitem [{\citenamefont {Gao}(2000)}]{gao00}%
  \BibitemOpen
  \bibfield  {author} {\bibinfo {author} {\bibfnamefont {B.}~\bibnamefont
  {Gao}},\ }\href@noop {} {\bibfield  {journal} {\bibinfo  {journal} {Phys.
  Rev. A}\ }\textbf {\bibinfo {volume} {62}},\ \bibinfo {pages} {050702(R)}
  (\bibinfo {year} {2000})}\BibitemShut {NoStop}%
\bibitem [{\citenamefont {Bartenstein}\ \emph {et~al.}(2005)\citenamefont
  {Bartenstein}, \citenamefont {Altmeyer}, \citenamefont {Riedl}, \citenamefont
  {Geusen}, \citenamefont {Jochim}, \citenamefont {Denschlag}, \citenamefont
  {Grimm}, \citenamefont {Simoni}, \citenamefont {Tiesinga}, \citenamefont
  {Williams},\ and\ \citenamefont {Julienne}}]{bar05}%
  \BibitemOpen
  \bibfield  {author} {\bibinfo {author} {\bibfnamefont {M.}~\bibnamefont
  {Bartenstein}}, \bibinfo {author} {\bibfnamefont {A.}~\bibnamefont
  {Altmeyer}}, \bibinfo {author} {\bibfnamefont {S.}~\bibnamefont {Riedl}},
  \bibinfo {author} {\bibfnamefont {R.}~\bibnamefont {Geusen}}, \bibinfo
  {author} {\bibfnamefont {S.}~\bibnamefont {Jochim}}, \bibinfo {author}
  {\bibfnamefont {C.~C. J.~H.}\ \bibnamefont {Denschlag}}, \bibinfo {author}
  {\bibfnamefont {R.}~\bibnamefont {Grimm}}, \bibinfo {author} {\bibfnamefont
  {A.}~\bibnamefont {Simoni}}, \bibinfo {author} {\bibfnamefont
  {E.}~\bibnamefont {Tiesinga}}, \bibinfo {author} {\bibfnamefont {C.~J.}\
  \bibnamefont {Williams}}, \ and\ \bibinfo {author} {\bibfnamefont {P.~S.}\
  \bibnamefont {Julienne}},\ }\href@noop {} {\bibfield  {journal} {\bibinfo
  {journal} {Phys. Rev. Lett.}\ }\textbf {\bibinfo {volume} {94}},\ \bibinfo
  {pages} {103201} (\bibinfo {year} {2005})}\BibitemShut {NoStop}%
\bibitem [{\citenamefont {Strecker}\ \emph {et~al.}(2003)\citenamefont
  {Strecker}, \citenamefont {Partridge},\ and\ \citenamefont {Hulet}}]{str03}%
  \BibitemOpen
  \bibfield  {author} {\bibinfo {author} {\bibfnamefont {K.~E.}\ \bibnamefont
  {Strecker}}, \bibinfo {author} {\bibfnamefont {G.~B.}\ \bibnamefont
  {Partridge}}, \ and\ \bibinfo {author} {\bibfnamefont {R.~G.}\ \bibnamefont
  {Hulet}},\ }\href@noop {} {\bibfield  {journal} {\bibinfo  {journal} {Phys.
  Rev. Lett.}\ }\textbf {\bibinfo {volume} {91}},\ \bibinfo {pages} {080406}
  (\bibinfo {year} {2003})}\BibitemShut {NoStop}%
\bibitem [{\citenamefont {Pollack}\ \emph {et~al.}(2009)\citenamefont
  {Pollack}, \citenamefont {Dries}, \citenamefont {Junker}, \citenamefont
  {Chen}, \citenamefont {Corcovilos},\ and\ \citenamefont {Hulet}}]{pol09}%
  \BibitemOpen
  \bibfield  {author} {\bibinfo {author} {\bibfnamefont {S.~E.}\ \bibnamefont
  {Pollack}}, \bibinfo {author} {\bibfnamefont {D.}~\bibnamefont {Dries}},
  \bibinfo {author} {\bibfnamefont {M.}~\bibnamefont {Junker}}, \bibinfo
  {author} {\bibfnamefont {Y.~P.}\ \bibnamefont {Chen}}, \bibinfo {author}
  {\bibfnamefont {T.~A.}\ \bibnamefont {Corcovilos}}, \ and\ \bibinfo {author}
  {\bibfnamefont {R.~G.}\ \bibnamefont {Hulet}},\ }\href@noop {} {\bibfield
  {journal} {\bibinfo  {journal} {Phys. Rev. Lett.}\ }\textbf {\bibinfo
  {volume} {102}},\ \bibinfo {eid} {090402} (\bibinfo {year}
  {2009})}\BibitemShut {NoStop}%
\bibitem [{\citenamefont {Strecker}\ \emph {et~al.}(2002)\citenamefont
  {Strecker}, \citenamefont {Partridge}, \citenamefont {Truscott},\ and\
  \citenamefont {Hulet}}]{str02}%
  \BibitemOpen
  \bibfield  {author} {\bibinfo {author} {\bibfnamefont {K.~E.}\ \bibnamefont
  {Strecker}}, \bibinfo {author} {\bibfnamefont {G.~B.}\ \bibnamefont
  {Partridge}}, \bibinfo {author} {\bibfnamefont {A.~G.}\ \bibnamefont
  {Truscott}}, \ and\ \bibinfo {author} {\bibfnamefont {R.~G.}\ \bibnamefont
  {Hulet}},\ }\href@noop {} {\bibfield  {journal} {\bibinfo  {journal}
  {Nature}\ }\textbf {\bibinfo {volume} {417}},\ \bibinfo {pages} {150}
  (\bibinfo {year} {2002})}\BibitemShut {NoStop}%
\bibitem [{\citenamefont {Inouye}\ \emph {et~al.}(1998)\citenamefont {Inouye},
  \citenamefont {Andrews}, \citenamefont {Stenger}, \citenamefont {Miesner},
  \citenamefont {Stamper-Kurn},\ and\ \citenamefont {Ketterle}}]{ino98}%
  \BibitemOpen
  \bibfield  {author} {\bibinfo {author} {\bibfnamefont {S.}~\bibnamefont
  {Inouye}}, \bibinfo {author} {\bibfnamefont {M.~R.}\ \bibnamefont {Andrews}},
  \bibinfo {author} {\bibfnamefont {J.}~\bibnamefont {Stenger}}, \bibinfo
  {author} {\bibfnamefont {H.-J.}\ \bibnamefont {Miesner}}, \bibinfo {author}
  {\bibfnamefont {D.~M.}\ \bibnamefont {Stamper-Kurn}}, \ and\ \bibinfo
  {author} {\bibfnamefont {W.}~\bibnamefont {Ketterle}},\ }\href@noop {}
  {\bibfield  {journal} {\bibinfo  {journal} {Nature}\ }\textbf {\bibinfo
  {volume} {392}},\ \bibinfo {pages} {151} (\bibinfo {year}
  {1998})}\BibitemShut {NoStop}%
\bibitem [{\citenamefont {Stenger}\ \emph {et~al.}(1999)\citenamefont
  {Stenger}, \citenamefont {Inouye}, \citenamefont {Andrews}, \citenamefont
  {Miesner}, \citenamefont {Stamper-Kurn},\ and\ \citenamefont
  {Ketterle}}]{ste99}%
  \BibitemOpen
  \bibfield  {author} {\bibinfo {author} {\bibfnamefont {J.}~\bibnamefont
  {Stenger}}, \bibinfo {author} {\bibfnamefont {S.}~\bibnamefont {Inouye}},
  \bibinfo {author} {\bibfnamefont {M.~R.}\ \bibnamefont {Andrews}}, \bibinfo
  {author} {\bibfnamefont {H.-J.}\ \bibnamefont {Miesner}}, \bibinfo {author}
  {\bibfnamefont {D.~M.}\ \bibnamefont {Stamper-Kurn}}, \ and\ \bibinfo
  {author} {\bibfnamefont {W.}~\bibnamefont {Ketterle}},\ }\href {\doibase
  10.1103/PhysRevLett.82.2422} {\bibfield  {journal} {\bibinfo  {journal}
  {Phys. Rev. Lett.}\ }\textbf {\bibinfo {volume} {82}},\ \bibinfo {pages}
  {2422} (\bibinfo {year} {1999})}\BibitemShut {NoStop}%
\bibitem [{\citenamefont {Regal}\ \emph {et~al.}(2004)\citenamefont {Regal},
  \citenamefont {Greiner},\ and\ \citenamefont {Jin}}]{reg04}%
  \BibitemOpen
  \bibfield  {author} {\bibinfo {author} {\bibfnamefont {C.~A.}\ \bibnamefont
  {Regal}}, \bibinfo {author} {\bibfnamefont {M.}~\bibnamefont {Greiner}}, \
  and\ \bibinfo {author} {\bibfnamefont {D.~S.}\ \bibnamefont {Jin}},\ }\href
  {\doibase 10.1103/PhysRevLett.92.040403} {\bibfield  {journal} {\bibinfo
  {journal} {Phys. Rev. Lett.}\ }\textbf {\bibinfo {volume} {92}},\ \bibinfo
  {pages} {040403} (\bibinfo {year} {2004})}\BibitemShut {NoStop}%
\bibitem [{\citenamefont {Regal}\ \emph {et~al.}(2003)\citenamefont {Regal},
  \citenamefont {Ticknor}, \citenamefont {Bohn},\ and\ \citenamefont
  {Jin}}]{reg03b}%
  \BibitemOpen
  \bibfield  {author} {\bibinfo {author} {\bibfnamefont {C.~A.}\ \bibnamefont
  {Regal}}, \bibinfo {author} {\bibfnamefont {C.}~\bibnamefont {Ticknor}},
  \bibinfo {author} {\bibfnamefont {J.~L.}\ \bibnamefont {Bohn}}, \ and\
  \bibinfo {author} {\bibfnamefont {D.~S.}\ \bibnamefont {Jin}},\ }\href@noop
  {} {\bibfield  {journal} {\bibinfo  {journal} {Phys. Rev. Lett.}\ }\textbf
  {\bibinfo {volume} {90}},\ \bibinfo {pages} {053201} (\bibinfo {year}
  {2003})}\BibitemShut {NoStop}%
\bibitem [{\citenamefont {Claussen}\ \emph {et~al.}(2003)\citenamefont
  {Claussen}, \citenamefont {Kokkelmans}, \citenamefont {Thompson},
  \citenamefont {Donley}, \citenamefont {Hodby},\ and\ \citenamefont
  {Wieman}}]{cla03}%
  \BibitemOpen
  \bibfield  {author} {\bibinfo {author} {\bibfnamefont {N.~R.}\ \bibnamefont
  {Claussen}}, \bibinfo {author} {\bibfnamefont {S.~J. J. M.~F.}\ \bibnamefont
  {Kokkelmans}}, \bibinfo {author} {\bibfnamefont {S.~T.}\ \bibnamefont
  {Thompson}}, \bibinfo {author} {\bibfnamefont {E.~A.}\ \bibnamefont
  {Donley}}, \bibinfo {author} {\bibfnamefont {E.}~\bibnamefont {Hodby}}, \
  and\ \bibinfo {author} {\bibfnamefont {C.~E.}\ \bibnamefont {Wieman}},\
  }\href@noop {} {\bibfield  {journal} {\bibinfo  {journal} {Phys. Rev. A}\
  }\textbf {\bibinfo {volume} {67}},\ \bibinfo {pages} {060701(R)} (\bibinfo
  {year} {2003})}\BibitemShut {NoStop}%
\bibitem [{\citenamefont {Volz}\ \emph {et~al.}(2003)\citenamefont {Volz},
  \citenamefont {D\"urr}, \citenamefont {Ernst}, \citenamefont {Marte},\ and\
  \citenamefont {Rempe}}]{vol03}%
  \BibitemOpen
  \bibfield  {author} {\bibinfo {author} {\bibfnamefont {T.}~\bibnamefont
  {Volz}}, \bibinfo {author} {\bibfnamefont {S.}~\bibnamefont {D\"urr}},
  \bibinfo {author} {\bibfnamefont {S.}~\bibnamefont {Ernst}}, \bibinfo
  {author} {\bibfnamefont {A.}~\bibnamefont {Marte}}, \ and\ \bibinfo {author}
  {\bibfnamefont {G.}~\bibnamefont {Rempe}},\ }\href {\doibase
  10.1103/PhysRevA.68.010702} {\bibfield  {journal} {\bibinfo  {journal} {Phys.
  Rev. A}\ }\textbf {\bibinfo {volume} {68}},\ \bibinfo {pages} {010702}
  (\bibinfo {year} {2003})}\BibitemShut {NoStop}%
\bibitem [{\citenamefont {D\"urr}\ \emph {et~al.}(2004)\citenamefont {D\"urr},
  \citenamefont {Volz},\ and\ \citenamefont {Rempe}}]{dur04}%
  \BibitemOpen
  \bibfield  {author} {\bibinfo {author} {\bibfnamefont {S.}~\bibnamefont
  {D\"urr}}, \bibinfo {author} {\bibfnamefont {T.}~\bibnamefont {Volz}}, \ and\
  \bibinfo {author} {\bibfnamefont {G.}~\bibnamefont {Rempe}},\ }\href
  {\doibase 10.1103/PhysRevA.70.031601} {\bibfield  {journal} {\bibinfo
  {journal} {Phys. Rev. A}\ }\textbf {\bibinfo {volume} {70}},\ \bibinfo
  {pages} {031601} (\bibinfo {year} {2004})}\BibitemShut {NoStop}%
\bibitem [{\citenamefont {Marte}\ \emph {et~al.}(2002)\citenamefont {Marte},
  \citenamefont {Volz}, \citenamefont {Schuster}, \citenamefont {{D\"{u}rr}},
  \citenamefont {Rempe}, \citenamefont {{van Kempen}},\ and\ \citenamefont
  {Verhaar}}]{mar02}%
  \BibitemOpen
  \bibfield  {author} {\bibinfo {author} {\bibfnamefont {A.}~\bibnamefont
  {Marte}}, \bibinfo {author} {\bibfnamefont {T.}~\bibnamefont {Volz}},
  \bibinfo {author} {\bibfnamefont {J.}~\bibnamefont {Schuster}}, \bibinfo
  {author} {\bibfnamefont {S.}~\bibnamefont {{D\"{u}rr}}}, \bibinfo {author}
  {\bibfnamefont {G.}~\bibnamefont {Rempe}}, \bibinfo {author} {\bibfnamefont
  {E.~G.~M.}\ \bibnamefont {{van Kempen}}}, \ and\ \bibinfo {author}
  {\bibfnamefont {B.}~\bibnamefont {Verhaar}},\ }\href@noop {} {\bibfield
  {journal} {\bibinfo  {journal} {Phys. Rev. Lett.}\ }\textbf {\bibinfo
  {volume} {89}},\ \bibinfo {pages} {283202} (\bibinfo {year}
  {2002})}\BibitemShut {NoStop}%
\bibitem [{\citenamefont {Widera}\ \emph {et~al.}(2004)\citenamefont {Widera},
  \citenamefont {Mandel}, \citenamefont {Greiner}, \citenamefont {Kreim},
  \citenamefont {H\"ansch},\ and\ \citenamefont {Bloch}}]{wid04}%
  \BibitemOpen
  \bibfield  {author} {\bibinfo {author} {\bibfnamefont {A.}~\bibnamefont
  {Widera}}, \bibinfo {author} {\bibfnamefont {O.}~\bibnamefont {Mandel}},
  \bibinfo {author} {\bibfnamefont {M.}~\bibnamefont {Greiner}}, \bibinfo
  {author} {\bibfnamefont {S.}~\bibnamefont {Kreim}}, \bibinfo {author}
  {\bibfnamefont {T.~W.}\ \bibnamefont {H\"ansch}}, \ and\ \bibinfo {author}
  {\bibfnamefont {I.}~\bibnamefont {Bloch}},\ }\href {\doibase
  10.1103/PhysRevLett.92.160406} {\bibfield  {journal} {\bibinfo  {journal}
  {Phys. Rev. Lett.}\ }\textbf {\bibinfo {volume} {92}},\ \bibinfo {pages}
  {160406} (\bibinfo {year} {2004})}\BibitemShut {NoStop}%
\bibitem [{\citenamefont {Chin}\ \emph {et~al.}(2004)\citenamefont {Chin},
  \citenamefont {Vuleti\ifmmode~\acute{c}\else \'{c}\fi{}}, \citenamefont
  {Kerman}, \citenamefont {Chu}, \citenamefont {Tiesinga}, \citenamefont
  {Leo},\ and\ \citenamefont {Williams}}]{chi04}%
  \BibitemOpen
  \bibfield  {author} {\bibinfo {author} {\bibfnamefont {C.}~\bibnamefont
  {Chin}}, \bibinfo {author} {\bibfnamefont {V.}~\bibnamefont
  {Vuleti\ifmmode~\acute{c}\else \'{c}\fi{}}}, \bibinfo {author} {\bibfnamefont
  {A.~J.}\ \bibnamefont {Kerman}}, \bibinfo {author} {\bibfnamefont
  {S.}~\bibnamefont {Chu}}, \bibinfo {author} {\bibfnamefont {E.}~\bibnamefont
  {Tiesinga}}, \bibinfo {author} {\bibfnamefont {P.~J.}\ \bibnamefont {Leo}}, \
  and\ \bibinfo {author} {\bibfnamefont {C.~J.}\ \bibnamefont {Williams}},\
  }\href {\doibase 10.1103/PhysRevA.70.032701} {\bibfield  {journal} {\bibinfo
  {journal} {Phys. Rev. A}\ }\textbf {\bibinfo {volume} {70}},\ \bibinfo
  {pages} {032701} (\bibinfo {year} {2004})}\BibitemShut {NoStop}%
\bibitem [{\citenamefont {Lange}\ \emph {et~al.}(2009)\citenamefont {Lange},
  \citenamefont {Pilch}, \citenamefont {Prantner}, \citenamefont {Ferlaino},
  \citenamefont {Engeser}, \citenamefont {Nagerl}, \citenamefont {Grimm},\ and\
  \citenamefont {Chin}}]{lan09}%
  \BibitemOpen
  \bibfield  {author} {\bibinfo {author} {\bibfnamefont {A.~D.}\ \bibnamefont
  {Lange}}, \bibinfo {author} {\bibfnamefont {K.}~\bibnamefont {Pilch}},
  \bibinfo {author} {\bibfnamefont {A.}~\bibnamefont {Prantner}}, \bibinfo
  {author} {\bibfnamefont {F.}~\bibnamefont {Ferlaino}}, \bibinfo {author}
  {\bibfnamefont {B.}~\bibnamefont {Engeser}}, \bibinfo {author} {\bibfnamefont
  {H.-C.}\ \bibnamefont {Nagerl}}, \bibinfo {author} {\bibfnamefont
  {R.}~\bibnamefont {Grimm}}, \ and\ \bibinfo {author} {\bibfnamefont
  {C.}~\bibnamefont {Chin}},\ }\href@noop {} {\bibfield  {journal} {\bibinfo
  {journal} {Phys. Rev. A}\ }\textbf {\bibinfo {volume} {79}},\ \bibinfo {eid}
  {013622} (\bibinfo {year} {2009})}\BibitemShut {NoStop}%
\bibitem [{\citenamefont {Chin}\ \emph {et~al.}(2000)\citenamefont {Chin},
  \citenamefont {{Vuleti\'{c}}}, \citenamefont {Kerman},\ and\ \citenamefont
  {Chu}}]{chi00}%
  \BibitemOpen
  \bibfield  {author} {\bibinfo {author} {\bibfnamefont {C.}~\bibnamefont
  {Chin}}, \bibinfo {author} {\bibfnamefont {V.}~\bibnamefont {{Vuleti\'{c}}}},
  \bibinfo {author} {\bibfnamefont {A.~J.}\ \bibnamefont {Kerman}}, \ and\
  \bibinfo {author} {\bibfnamefont {S.}~\bibnamefont {Chu}},\ }\href@noop {}
  {\bibfield  {journal} {\bibinfo  {journal} {Phys. Rev. Lett.}\ }\textbf
  {\bibinfo {volume} {85}},\ \bibinfo {pages} {2717} (\bibinfo {year}
  {2000})}\BibitemShut {NoStop}%
\bibitem [{\citenamefont {Ticknor}\ \emph {et~al.}(2004)\citenamefont
  {Ticknor}, \citenamefont {Regal}, \citenamefont {Jin},\ and\ \citenamefont
  {Bohn}}]{tic04}%
  \BibitemOpen
  \bibfield  {author} {\bibinfo {author} {\bibfnamefont {C.}~\bibnamefont
  {Ticknor}}, \bibinfo {author} {\bibfnamefont {C.~A.}\ \bibnamefont {Regal}},
  \bibinfo {author} {\bibfnamefont {D.~S.}\ \bibnamefont {Jin}}, \ and\
  \bibinfo {author} {\bibfnamefont {J.~L.}\ \bibnamefont {Bohn}},\ }\href@noop
  {} {\bibfield  {journal} {\bibinfo  {journal} {Phys. Rev. A}\ }\textbf
  {\bibinfo {volume} {69}},\ \bibinfo {pages} {042712} (\bibinfo {year}
  {2004})}\BibitemShut {NoStop}%
\bibitem [{\citenamefont {Zhang}\ \emph {et~al.}(2004)\citenamefont {Zhang},
  \citenamefont {van Kempen}, \citenamefont {Bourdel}, \citenamefont
  {Khaykovich}, \citenamefont {Cubizolles}, \citenamefont {Chevy},
  \citenamefont {Teichmann}, \citenamefont {Tarruell}, \citenamefont
  {Kokkelmans},\ and\ \citenamefont {Salomon}}]{zha04}%
  \BibitemOpen
  \bibfield  {author} {\bibinfo {author} {\bibfnamefont {J.}~\bibnamefont
  {Zhang}}, \bibinfo {author} {\bibfnamefont {E.~G.~M.}\ \bibnamefont {van
  Kempen}}, \bibinfo {author} {\bibfnamefont {T.}~\bibnamefont {Bourdel}},
  \bibinfo {author} {\bibfnamefont {L.}~\bibnamefont {Khaykovich}}, \bibinfo
  {author} {\bibfnamefont {J.}~\bibnamefont {Cubizolles}}, \bibinfo {author}
  {\bibfnamefont {F.}~\bibnamefont {Chevy}}, \bibinfo {author} {\bibfnamefont
  {M.}~\bibnamefont {Teichmann}}, \bibinfo {author} {\bibfnamefont
  {L.}~\bibnamefont {Tarruell}}, \bibinfo {author} {\bibfnamefont {S.~J. J.
  M.~F.}\ \bibnamefont {Kokkelmans}}, \ and\ \bibinfo {author} {\bibfnamefont
  {C.}~\bibnamefont {Salomon}},\ }\href {\doibase 10.1103/PhysRevA.70.030702}
  {\bibfield  {journal} {\bibinfo  {journal} {Phys. Rev. A}\ }\textbf {\bibinfo
  {volume} {70}},\ \bibinfo {pages} {030702} (\bibinfo {year}
  {2004})}\BibitemShut {NoStop}%
\bibitem [{\citenamefont {Schunck}\ \emph {et~al.}(2005)\citenamefont
  {Schunck}, \citenamefont {Zwierlein}, \citenamefont {Stan}, \citenamefont
  {Raupach}, \citenamefont {Ketterle}, \citenamefont {Simoni}, \citenamefont
  {Tiesinga}, \citenamefont {Williams},\ and\ \citenamefont
  {Julienne}}]{sch05}%
  \BibitemOpen
  \bibfield  {author} {\bibinfo {author} {\bibfnamefont {C.~H.}\ \bibnamefont
  {Schunck}}, \bibinfo {author} {\bibfnamefont {M.~W.}\ \bibnamefont
  {Zwierlein}}, \bibinfo {author} {\bibfnamefont {C.~A.}\ \bibnamefont {Stan}},
  \bibinfo {author} {\bibfnamefont {S.~M.~F.}\ \bibnamefont {Raupach}},
  \bibinfo {author} {\bibfnamefont {W.}~\bibnamefont {Ketterle}}, \bibinfo
  {author} {\bibfnamefont {A.}~\bibnamefont {Simoni}}, \bibinfo {author}
  {\bibfnamefont {E.}~\bibnamefont {Tiesinga}}, \bibinfo {author}
  {\bibfnamefont {C.~J.}\ \bibnamefont {Williams}}, \ and\ \bibinfo {author}
  {\bibfnamefont {P.~S.}\ \bibnamefont {Julienne}},\ }\href@noop {} {\bibfield
  {journal} {\bibinfo  {journal} {Phys. Rev. A}\ }\textbf {\bibinfo {volume}
  {71}},\ \bibinfo {pages} {045601} (\bibinfo {year} {2005})}\BibitemShut
  {NoStop}%
\bibitem [{\citenamefont {Gaebler}\ \emph {et~al.}(2007)\citenamefont
  {Gaebler}, \citenamefont {Stewart}, \citenamefont {Bohn},\ and\ \citenamefont
  {Jin}}]{gae07}%
  \BibitemOpen
  \bibfield  {author} {\bibinfo {author} {\bibfnamefont {J.~P.}\ \bibnamefont
  {Gaebler}}, \bibinfo {author} {\bibfnamefont {J.~T.}\ \bibnamefont
  {Stewart}}, \bibinfo {author} {\bibfnamefont {J.~L.}\ \bibnamefont {Bohn}}, \
  and\ \bibinfo {author} {\bibfnamefont {D.~S.}\ \bibnamefont {Jin}},\
  }\href@noop {} {\bibfield  {journal} {\bibinfo  {journal} {Phys. Rev. Lett.}\
  }\textbf {\bibinfo {volume} {98}},\ \bibinfo {pages} {200403} (\bibinfo
  {year} {2007})}\BibitemShut {NoStop}%
\bibitem [{\citenamefont {Knoop}\ \emph {et~al.}(2008)\citenamefont {Knoop},
  \citenamefont {Mark}, \citenamefont {Ferlaino}, \citenamefont {Danzl},
  \citenamefont {Kraemer}, \citenamefont {N\"{a}gerl},\ and\ \citenamefont
  {Grimm}}]{kno08}%
  \BibitemOpen
  \bibfield  {author} {\bibinfo {author} {\bibfnamefont {S.}~\bibnamefont
  {Knoop}}, \bibinfo {author} {\bibfnamefont {M.}~\bibnamefont {Mark}},
  \bibinfo {author} {\bibfnamefont {F.}~\bibnamefont {Ferlaino}}, \bibinfo
  {author} {\bibfnamefont {J.~G.}\ \bibnamefont {Danzl}}, \bibinfo {author}
  {\bibfnamefont {T.}~\bibnamefont {Kraemer}}, \bibinfo {author} {\bibfnamefont
  {H.-C.}\ \bibnamefont {N\"{a}gerl}}, \ and\ \bibinfo {author} {\bibfnamefont
  {R.}~\bibnamefont {Grimm}},\ }\href {\doibase 10.1103/PhysRevLett.100.083002}
  {\bibfield  {journal} {\bibinfo  {journal} {Phys. Rev. Lett.}\ }\textbf
  {\bibinfo {volume} {100}},\ \bibinfo {eid} {083002} (\bibinfo {year}
  {2008})}\BibitemShut {NoStop}%
\bibitem [{\citenamefont {Fuchs}\ \emph {et~al.}(2008)\citenamefont {Fuchs},
  \citenamefont {Ticknor}, \citenamefont {Dyke}, \citenamefont {Veeravalli},
  \citenamefont {Kuhnle}, \citenamefont {Rowlands}, \citenamefont {Hannaford},\
  and\ \citenamefont {Vale}}]{fuc08}%
  \BibitemOpen
  \bibfield  {author} {\bibinfo {author} {\bibfnamefont {J.}~\bibnamefont
  {Fuchs}}, \bibinfo {author} {\bibfnamefont {C.}~\bibnamefont {Ticknor}},
  \bibinfo {author} {\bibfnamefont {P.}~\bibnamefont {Dyke}}, \bibinfo {author}
  {\bibfnamefont {G.}~\bibnamefont {Veeravalli}}, \bibinfo {author}
  {\bibfnamefont {E.}~\bibnamefont {Kuhnle}}, \bibinfo {author} {\bibfnamefont
  {W.}~\bibnamefont {Rowlands}}, \bibinfo {author} {\bibfnamefont
  {P.}~\bibnamefont {Hannaford}}, \ and\ \bibinfo {author} {\bibfnamefont
  {C.~J.}\ \bibnamefont {Vale}},\ }\href {\doibase 10.1103/PhysRevA.77.053616}
  {\bibfield  {journal} {\bibinfo  {journal} {Phys. Rev. A}\ }\textbf {\bibinfo
  {volume} {77}},\ \bibinfo {pages} {053616} (\bibinfo {year}
  {2008})}\BibitemShut {NoStop}%
\bibitem [{\citenamefont {Lysebo}\ and\ \citenamefont {Veseth}(2009)}]{lys09}%
  \BibitemOpen
  \bibfield  {author} {\bibinfo {author} {\bibfnamefont {M.}~\bibnamefont
  {Lysebo}}\ and\ \bibinfo {author} {\bibfnamefont {L.}~\bibnamefont
  {Veseth}},\ }\href {\doibase 10.1103/PhysRevA.79.062704} {\bibfield
  {journal} {\bibinfo  {journal} {Phys. Rev. A}\ }\textbf {\bibinfo {volume}
  {79}},\ \bibinfo {pages} {062704} (\bibinfo {year} {2009})}\BibitemShut
  {NoStop}%
\bibitem [{\citenamefont {Landau}\ and\ \citenamefont
  {Lifshitz}(1977)}]{lan77}%
  \BibitemOpen
  \bibfield  {author} {\bibinfo {author} {\bibfnamefont {L.~D.}\ \bibnamefont
  {Landau}}\ and\ \bibinfo {author} {\bibfnamefont {E.~M.}\ \bibnamefont
  {Lifshitz}},\ }\href@noop {} {\emph {\bibinfo {title} {Quantum Mechanics}}}\
  (\bibinfo  {publisher} {Pergamon Press, Oxford},\ \bibinfo {year}
  {1977})\BibitemShut {NoStop}%
\bibitem [{\citenamefont {Bruun}\ \emph {et~al.}(2005)\citenamefont {Bruun},
  \citenamefont {Jackson},\ and\ \citenamefont {Kolomeitsev}}]{bru05}%
  \BibitemOpen
  \bibfield  {author} {\bibinfo {author} {\bibfnamefont {G.~M.}\ \bibnamefont
  {Bruun}}, \bibinfo {author} {\bibfnamefont {A.~D.}\ \bibnamefont {Jackson}},
  \ and\ \bibinfo {author} {\bibfnamefont {E.~E.}\ \bibnamefont
  {Kolomeitsev}},\ }\href {\doibase 10.1103/PhysRevA.71.052713} {\bibfield
  {journal} {\bibinfo  {journal} {Phys. Rev. A}\ }\textbf {\bibinfo {volume}
  {71}},\ \bibinfo {pages} {052713} (\bibinfo {year} {2005})}\BibitemShut
  {NoStop}%
\bibitem [{\citenamefont {Marcelis}\ \emph {et~al.}(2004)\citenamefont
  {Marcelis}, \citenamefont {van Kempen}, \citenamefont {Verhaar},\ and\
  \citenamefont {Kokkelmans}}]{mar04}%
  \BibitemOpen
  \bibfield  {author} {\bibinfo {author} {\bibfnamefont {B.}~\bibnamefont
  {Marcelis}}, \bibinfo {author} {\bibfnamefont {E.~G.~M.}\ \bibnamefont {van
  Kempen}}, \bibinfo {author} {\bibfnamefont {B.~J.}\ \bibnamefont {Verhaar}},
  \ and\ \bibinfo {author} {\bibfnamefont {S.~J. J. M.~F.}\ \bibnamefont
  {Kokkelmans}},\ }\href@noop {} {\bibfield  {journal} {\bibinfo  {journal}
  {Phys. Rev. A}\ }\textbf {\bibinfo {volume} {70}},\ \bibinfo {pages} {012701}
  (\bibinfo {year} {2004})}\BibitemShut {NoStop}%
\end{thebibliography}%

\end{document}